\documentclass[12pt,onecolumn,draftcls]{IEEEtran}
\usepackage[latin9]{inputenc}
\usepackage{geometry}
\usepackage{xcolor}
\usepackage{textcomp}
\usepackage{amsmath}
\usepackage{amssymb}
\usepackage{graphicx}
\usepackage{setspace}

\makeatletter

\providecommand{\tabularnewline}{\\}

\let\oldforeign@language\foreign@language
\DeclareRobustCommand{\foreign@language}[1]{%
  \lowercase{\oldforeign@language{#1}}}

\usepackage{color}
\usepackage{textcomp}
\usepackage{scalefnt}
\usepackage{multicol}
\usepackage[font=small,skip=2pt]{caption}

\providecommand{\tabularnewline}{\\}
\let\oldforeign@language\foreign@language
\DeclareRobustCommand{\foreign@language}[1]{  \lowercase{\oldforeign@language{#1}}}

\makeatother

\begin{document}

\title{Direct Data Detection of OFDM Signals Over Wireless Channels}

\author{A. Saci,~\IEEEmembership{Student Member,~IEEE}, A. Al-Dweik,~\IEEEmembership{Senior Member,~IEEE},
A. Shami,~\IEEEmembership{Senior Member,~IEEE}\thanks{A. Saci, A. Al-Dweik and A. Shami are with the Department of Electrical
and Computer Engineering, Western University, London, ON, Canada,
(e-mail: \{asaci, aaldweik, abdallah.shami\}@uwo.ca).}\thanks{A. Al-Dweik is also with the Department of Electrical and Computer
Engineering, Khalifa University, Abu Dhabi, UAE, (e-mail: dweik@kustar.ac.ae).}\thanks{Part of this work is protected by the US patent: A. Al-Dweik ``Signal
detection in a communication system.'' U.S. Patent No. 9,596,119.
14 Mar. 2017.}}
\maketitle
\begin{abstract}
This paper presents a novel efficient receiver design for wireless
communication systems that incorporate orthogonal frequency division
multiplexing (OFDM) transmission. The proposed receiver does not require
channel estimation or equalization to perform coherent data detection.
Instead, channel estimation, equalization, and data detection are
combined into a single operation, and hence, the detector is denoted
as a direct data detector ($D^{3}$). The performance of the proposed
system is thoroughly analyzed theoretically in terms of bit error
rate (BER), and validated by Monte Carlo simulations. The obtained
theoretical and simulation results demonstrate that the BER of the
proposed $D^{3}$ is only $3$ dB away from coherent detectors with
perfect knowledge of the channel state information (CSI) in flat fading
channels, and similarly in frequency-selective channels for a wide
range of signal-to-noise ratios (SNRs). If CSI is not known perfectly,
then the $D^{3}$ outperforms the coherent detector substantially,
particularly at high SNRs with linear interpolation. The computational
complexity of the $D^{3}$ depends on the length of the sequence to
be detected, nevertheless, a significant complexity reduction can
be achieved using the Viterbi algorithm. 
\end{abstract}

\markboth{IEEE Transactions on Vehicular Technology}{Shell \MakeLowercase{\textit{et al.}}: Bare Demo of IEEEtran.cls
for IEEE Journals}

\section{Introduction}

\textcolor{black}{Orthogonal frequency division multiplexing (OFDM)
is widely adopted in several wired and wireless communication standards,
such as worldwide interoperability for microwave access (WiMAX) technologies
\cite{WiMax}, Long Term Evolution-Advanced (LTE-A) standard \cite{LTE-A},
Digital Video Broadcasting (DVB), Terrestrial (DVB-T) and Hand-held
(DVB-H) \cite{DVB-T}, optical wireless communications (OWC) \cite{Tsonev-00},
\cite{Armstrong-01}, and recently, it has been adopted for the fifth-generation
(5G) wireless networks \cite{5G}. The channel is typically modeled
as frequency-selective for WiMax and LTE-A, flat for OWC in the presence
of atmospheric turbulence \cite{Tsonev-00}, \cite{Armstrong-01}.
Therefore, OFDM has become the lead above other modulation schemes
at present and in the near future \cite{5G-survey}.}

\textcolor{black}{One of the main advantages of OFDM is that each
subcarrier experiences flat fading even though the overall signal
spectrum suffers from frequency-selective fading.}\textcolor{green}{{}
}\textcolor{black}{Moreover, incorporating the concept of a cyclic
prefix (CP), which is formed by copying a part of the OFDM symbol
of and pre-append it to the transmitted OFDM block, prevents intersymbol
interference (ISI) if the CP length is larger than the maximum delay
spread of the channel.} Consequently, a low-complexity single-tap
equalizer can be utilized to eliminate the impact of the multipath
fading channel. Under such circumstances, the OFDM demodulation process
can be performed once the fading parameters at each subcarrier, commonly
denoted as channel state information (CSI), are estimated.

In general, channel estimation can be classified into blind \cite{One-Shot-CFO-2014}-\cite{blind-massive-mimo-acd},
and pilot-aided techniques \cite{Robust-CE-OFDM-2015}-\cite{pilot-ce-pilot-freq-domain}.
Blind channel estimation techniques are spectrally efficient because
they do not require any overhead to estimate the CSI, nevertheless,
such techniques have not yet been adopted in practical OFDM systems.
Conversely, pilot-based CSI estimation is preferred for practical
systems, because typically it is more robust and less complex. In
pilot-based CSI estimation, the pilot symbols are embedded within
the subcarriers of the transmitted OFDM signal in time and frequency
domain; hence, the pilots form a two dimensional (2-D) grid \cite{LTE-A}.
The channel response at the pilot symbols can be obtained using the
least-squares (LS) frequency domain estimation, and the channel parameters
at other subcarriers can be obtained using various interpolation techniques
\cite{Rayleigh-Ricean-Interpolation-TCOM2008}. Optimal interpolation
requires a 2-D Wiener filter that exploits the time and frequency
correlation of the channel, however, it is substantially complex to
implement \cite{Interpolation-TCOM-2010}, \cite{Wiener}. The complexity
can be reduced by decomposing the 2-D interpolation process into two
cascaded 1-D processes, and then, using less computationally-involved
interpolation schemes \cite{Adaptive-Equalization-IEEE-Broadcasting-2008},
\cite{Comp-Pilot-VTC2007}. Low complexity interpolation, however,
is usually accompanied by error rate performance degradation \cite{Comp-Pilot-VTC2007}.
It is also worth noting that most practical OFDM-based systems utilize
a fixed grid pattern structure \cite{LTE-A}.

Once the channel parameters are obtained for all subcarriers, the
received samples at the output of the fast Fourier transform (FFT)
are equalized to compensate for the channel fading. Fortunately, the
equalization for OFDM is performed in the frequency domain using single-tap
equalizers. The equalizer output samples, which are denoted as the
decision variables, will be applied to a maximum likelihood detector
(MLD) to regenerate the information symbols.

In addition to the direct approach, several techniques have been proposed
in the literature to estimate the CSI or detect the data symbols indirectly,
by exploiting the correlation among the channel coefficients. For
example, the per-survivor processing (PSP) approach has been widely
used to approximate the maximum likelihood sequence estimator (MLSE)
for coded and uncoded sequences \cite{PSP-Raheli}, \cite{PSP-Zhu},
\cite{Rev-1}. The PSP utilizes the Viterbi algorithm (VA) to recursively
estimate the CSI without interpolation using the least mean squares
(LMS) algorithm. Although the PSP provides superior performance when
the channel is flat over the entire sequence, its performance degrades
severely if this condition is not satisfied, even when the LMS step
size is adaptive \cite{PSP-Zhu}. Multiple symbol differential detection
(MSDD) can be also used for sequence estimation without explicit channel
estimation. In such systems, the information is embedded in the phase
difference between adjacent symbols, and hence, differential encoding
is needed. Although differential detection is only $3$ dB worse than
coherent detection in flat fading channels, its performance may deteriorate
significantly in frequency-selective channels \cite{Divsalar}, \cite{Diff-Xhang}.
Consequently, Wu and Kam \cite{Wu 2010} proposed a generalized likelihood
ratio test (GLRT) receiver whose performance without CSI is comparable
to the coherent detector. Although the GLRT receiver is more robust
than differential detectors in frequency-selective channels, its performance
is significantly worse than coherent detectors.

\textcolor{black}{The estimator-correlator (EC) cross-correlates the
received signal with an estimate of the channel output signal corresponding
to each possible transmitted signal \cite{Estimator_Correlator_2},
\cite{Estimator-Correlator}. }The signal at channel output is estimated
with a minimum mean square error (MMSE) estimator from the knowledge
of the received signal and the second order statistics of the channel
and no\textcolor{black}{ise. The channel estimation (CE) m}ay provide
BER that is about $1$ dB from the ML coherent detector in flat fading
channels but at the expense of a large number of pilots. \textcolor{black}{Moreover,
the BER performance of EC detectors is generally poor in frequency-selective
channels where the CE BER is significantly worse than the ML coherent
detector \cite{Estimator-Correlator}.} Decision-directed techniques
can also be used to avoid conventional channel estimation. For example,
the authors in \cite{Saci-Tcom} proposed a hybrid frame structure
that enables blind decision-directed channel estimation. Although
the proposed system manages to offer reliable channel estimates and
BER in various channel conditions, the system structure follows the
typical coherent detector design where equalization and symbol detection
are required.

\subsection{Motivation and Key Contributions}

Unlike conventional OFDM detectors, this work presents a new detector
to regenerate the information symbols directly from the received samples
at the FFT output, which is denoted as the direct data detector ($D^{3})$.
By using the $D^{3}$, there is no need to perform channel estimation,
interpolation, equalization, or symbol decision operations. The $D^{3}$
exploits the fact that channel coefficients over adjacent subcarriers
are highly correlated and approximately equal. Consequently, the $D^{3}$
is derived by minimizing the difference between channel coefficients
of adjacent subcarriers. The main limitation of the $D^{3}$ is that
it suffers from a phase ambiguity problem, which can be solved using
pilot symbols, which are part of a transmission frame in most practical
standards \cite{WiMax}, \cite{LTE-A}. To the best of the authors'
knowledge, there is no work reported in the published literature that
uses the proposed principle.

The $D^{3}$ performance is evaluated in terms of complexity, computational
power, and bit error rate (BER), where analytic expressions are derived
for several channel models and system configurations. The $D^{3}$
BER is compared to other widely used detectors such as the maximum
likelihood (ML) coherent detector \cite{Proakis-Book-2001} with perfect
and imperfect CSI, multiple symbol differential detector (MSDD) \cite{Divsalar},
the ML sequence detector (MLSD) with no CSI \cite{Wu 2010}, and the
per-survivor processing detector \cite{PSP-Raheli}. The obtained
results show that the $D^{3}$ is more robust than all the other considered
detectors in various cases of interest, particularly in frequency-selective
channels at moderate and high SNRs. Moreover, the computational power
comparison shows that the $D^{3}$ requires less than $35\%$ of the
computational power required by the ML coherent detector.

\subsection{Paper Organization and Notations}

The rest of this paper is organized as follows. The OFDM system and
channel models are described in Section \ref{sec:Signal-and-Channel}.
The proposed $D^{3}$ is presented in Section \ref{sec:Proposed-System-Model},
and the efficient implementation of the $D^{3}$ is explored in Section
\ref{sec:Efficient-Implementation-of-D3}. The system error probability
performance analysis is presented in Section \ref{sec:System-Performance-Analysis}.
Complexity analysis of the conventional pilot based OFDM and the $D^{3}$
are given in Section \ref{sec:Complexity-Analysis}. Numerical results
are discussed in Section \ref{sec:Numerical-Results}, and finally,
the conclusion is drawn in Section \ref{sec:Conclusion}.

In what follows, unless otherwise specified, uppercase boldface and
blackboard letters such as $\mathbf{H}$ and $\mathbb{H}$, will denote
$N\times N$ matrices, whereas lowercase boldface letters such as
$\mathbf{x}$ will denote row or column vectors with $N$ elements.
Uppercase, lowercase, or bold letters with a tilde such as $\tilde{d}$
will denote trial values, and symbols with a hat, such as $\hat{\mathbf{x}}$,
will denote the estimate of $\mathbf{x}$. Letters with apostrophe
such as $\acute{v}$ are used to denote the next value, i.e., $\acute{v}\triangleq v+1$.
Furthermore, $\mathrm{E}\left[\cdot\right]$ denotes the expectation
operation.

\section{Signal and Channel Models \label{sec:Signal-and-Channel}}

Consider an OFDM system with $N$ subcarriers modulated by a sequence
of $N$ complex data symbols $\mathbf{d}=[d_{0}$, $d_{1}$, $....$,
$d_{N-1}]^{T}$. The data symbols are selected uniformly from a general
constellation such as $M$-ary phase shift keying (MPSK) or quadrature
amplitude modulation (QAM). In conventional pilot-aided OFDM systems
\cite{IEEE-AC}, $N_{P}$ of the subcarriers are allocated for pilot
symbols, which can be used for channel estimation and synchronization
purposes. The modulation process in OFDM can be implemented efficiently
using an $N$-point inverse FFT (IFFT) algorithm, where its output
during the $\ell$th OFDM block can be written as $\mathbf{x(\ell)=F}^{H}\mathbf{d(\ell)}$
where $\mathbf{F}$ is the normalized $N\times N$ FFT matrix, and
hence, $\mathbf{F}^{H}$ is the IFFT matrix. To simplify the notation,
the block index $\ell$ is dropped for the remaining parts of the
paper unless it is necessary to include it. Then, a CP of length $N_{\mathrm{CP}}$
samples, no less than the channel maximum delay spread ($\mathcal{D}_{\mathrm{h}}$),
is appended to compose the OFDM symbol with a total length $N_{\mathrm{t}}=N+N_{\mathrm{CP}}$
samples and duration of $T_{\mathrm{t}}$s.

At the receiver front-end, the received signal is down-converted to
baseband and sampled at a rate $T_{\mathrm{s}}=T_{\mathrm{t}}/N_{\mathrm{t}}$.
In this work, the channel is assumed to be composed of $\mathcal{D}_{\mathrm{h}}+1$
independent multipath components each of which has a gain $h_{m}\sim\mathcal{CN}\left(0,2\sigma_{h_{m}}^{2}\right)$
and delay $m\times T_{\mathrm{s}}$, where $m\in\{0$, $1$,$...$,
$\mathcal{D}_{\mathrm{h}}\}$. A quasi-static channel is assumed throughout
this work, and thus, the channel taps are considered constant over
one OFDM symbol, but they may change over two consecutive symbols.
Therefore, the received sequence after dropping the CP samples and
applying the FFT can be expressed as, 
\begin{equation}
\mathbf{r}=\mathbf{Hd+w}\label{eq:rx_sig_FD}
\end{equation}
where $\left\{ \mathbf{r,w}\right\} \in\mathbb{C}^{N\times1}$, $w_{v}\sim\mathcal{CN}\left(0\text{, }2\sigma_{w}^{2}\right)$
is the additive white Gaussian noise (AWGN) vector and $\mathbf{\mathbf{H}}$
denotes the channel frequency response (CFR) 
\begin{equation}
\mathbf{\mathbf{H}}=\text{diag}\left\{ \left[H_{0}\text{, }H_{1}\text{,}\ldots\text{, }H_{N-1}\right]\right\} .
\end{equation}
By noting that $\mathbf{r|}_{\mathbf{H,d}}\sim\mathcal{CN}\left(\mathbf{Hd}\text{, }2\sigma_{w}^{2}\mathbf{I}_{N}\right)$
where $\mathbf{I}_{N}$ is an $N\times N$ identity matrix, then it
is straightforward to show that the MLD can be expressed as 
\begin{equation}
\mathbf{\hat{d}}=\arg\text{ }\min_{\tilde{\mathbf{d}}}\text{ }\left\Vert \mathbf{r-H}\tilde{\mathbf{d}}\right\Vert ^{2}\label{E-MLD-01}
\end{equation}
where $\left\Vert \mathbf{\cdot}\right\Vert $ denotes the Euclidean
norm, and $\tilde{\mathbf{d}}=\left[\tilde{d}_{0}\text{, }\tilde{d}_{1}\text{,}\ldots\text{, }\tilde{d}_{N1}\right]^{T}$
denotes the trial values of $\mathbf{d}$. As can be noted from (\ref{E-MLD-01}),
the MLD requires the knowledge of $\mathbf{\mathbf{H}}$. Moreover,
because (\ref{E-MLD-01}) describes the detection of more than one
symbol, it is typically denoted as maximum likelihood sequence detector
(MLSD). If the elements of $\mathbf{d}$ are independent, the MLSD
can be replaced by a symbol-by-symbol MLD 
\begin{equation}
\hat{d}_{v}=\arg\text{ }\min_{\tilde{d}_{v}}\text{ }\left\vert r_{v}\mathbf{-}H_{v}\tilde{d}_{v}\right\vert ^{2}\text{.}\label{E-MLD-02}
\end{equation}
Since perfect knowledge of $\mathbf{H}$ is infeasible, an estimated
version of $\mathbf{H}$, denoted as $\hat{\mathbf{H}}$, can be used
in (\ref{E-MLD-01}) and (\ref{E-MLD-02}) instead of $\mathbf{H}$\textbf{.
}Another possible approach to implement the detector is to equalize
$\mathbf{r}$, and then use a symbol-by-symbol MLD. \textcolor{black}{Because
the considered system is assumed to have no ISI or intercarrier interference
(ICI), then a single-tap frequency-domain zero-forcing equalizer can
be used. }Therefore, the equalized received sequence can be expressed
as, 
\begin{equation}
\check{\mathbf{r}}=\left[\hat{\mathbf{H}}^{H}\hat{\mathbf{H}}\right]^{-1}\hat{\mathbf{H}}^{H}\mathbf{r}
\end{equation}
and 
\begin{equation}
\hat{d}_{v}=\arg\min_{\tilde{d}_{v}}\left\vert \check{r}_{v}-\tilde{d}_{v}\right\vert ^{2}\text{, }\forall v\text{.}
\end{equation}

It is interesting to note that solving (\ref{E-MLD-01}) does not
necessarily require the explicit knowledge of $\mathbf{H}$ under
some special circumstances. For example, Wu and Kam \cite{Wu 2010}
noticed that in flat fading channels, i.e., $H_{v}=H$ $\forall v$,
it is possible to detect the data symbols using the following MLSD,
\begin{equation}
\mathbf{\hat{d}}=\arg\text{ }\max_{\tilde{\mathbf{d}}}\text{ }\frac{\left\vert \tilde{\mathbf{d}}^{H}\mathbf{r}\right\vert ^{2}}{\parallel\tilde{\mathbf{d}}\Vert}.\label{E-Wu}
\end{equation}
Although the detector described in (\ref{E-Wu}) is efficient in the
sense that it does not require the knowledge of $\mathbf{H}$, its
BER is very sensitive to the channel variations.

\section{Proposed $D^{3}$ System Model\label{sec:Proposed-System-Model}}

One of the distinctive features of OFDM is that its channel coefficients
over adjacent subcarriers in the frequency domain are highly correlated
and approximately equal. The correlation coefficient between two adjacent
subcarriers can be defined as 
\begin{eqnarray}
\varrho_{f} & \triangleq & \mathrm{E}\left[H_{v}H_{\acute{v}}^{\ast}\right]\nonumber \\
 & = & \mathrm{E}\left[\sum_{n=0}^{\mathcal{D}_{\mathrm{h}}}h_{n}e^{-j2\pi\frac{nv}{N}}\sum_{m=0}^{\mathcal{D}_{\mathrm{h}}}h_{m}^{\ast}e^{j2\pi\frac{m\acute{v}}{N}}\right]=\sum_{m=0}^{\mathcal{D}_{\mathrm{h}}}\sigma_{h_{m}}^{2}e^{j2\pi\frac{m}{N}}\label{eq:rho-f}
\end{eqnarray}
where $\sigma_{h_{m}}^{2}=\mathrm{E}\left[\left\vert h_{m}\right\vert ^{2}\right]$.
The difference between two adjacent channel coefficients is 
\begin{equation}
\Delta_{f}=\mathrm{E}\left[H_{v}-H_{\acute{v}}\right]=\mathrm{E}\left[\sum_{m=0}^{\mathcal{D}_{\mathrm{h}}}h_{n}e^{-j2\pi\frac{mv}{N}}\left(1-e^{-j2\pi\frac{m}{N}}\right)\right]
\end{equation}
For large values of $N$, it is straightforward to show that $\varrho_{f}\rightarrow1$
and $\Delta_{f}\rightarrow0$. Similar to the frequency domain, the
time domain correlation defined according to the Jakes' model can
be computed as \cite{Jakes-Model}, 
\begin{equation}
\varrho_{t}=\mathrm{E}\left[H_{v}^{\ell}\left(H_{v}^{\acute{\ell}}\right)^{\ast}\right]=J_{0}\left(2\pi f_{d}T_{\mathrm{s}}\right)\label{eq:rho-t}
\end{equation}
where $J_{0}\left(\cdot\right)$ is the Bessel function of the first
kind and $0$ order, $f_{d}$ is the maximum Doppler frequency. For
large values of $N$, $2\pi f_{d}T_{\mathrm{s}}\ll1$, and hence $J_{0}\left(2\pi f_{d}T_{\mathrm{s}}\right)\approx1$,
and thus $\varrho_{t}\approx1$. Using the same argument, the difference
in the time domain $\Delta_{t}\triangleq\mathrm{E}\left[H_{v}^{\ell}-H_{v}^{\acute{\ell}}\right]\approx0$.
Although the proposed system can be applied in the time domain, frequency
domain, or both, the focus of this work is the frequency domain.

Based on the aforementioned properties of OFDM, a simple approach
to extract the information symbols from the received sequence $\mathbf{r}$
can be designed by minimizing the difference of the channel coefficients
between adjacent subcarriers, which can be expressed as 
\begin{equation}
\mathbf{\hat{d}}=\arg\min_{\tilde{\mathbf{d}}}\sum_{v=0}^{N-2}\left\vert \frac{r_{v}}{\tilde{d}_{v}}-\frac{r_{\acute{v}}}{\tilde{d}_{\acute{v}}}\right\vert ^{2}.\label{E-DDD-00}
\end{equation}
As can be noted from (\ref{E-DDD-00}), the estimated data sequence
$\mathbf{\hat{d}}$ can be obtained without the knowledge of $\mathbf{H}$.
Moreover, there is no requirement for the channel coefficients over
the considered sequence to be equal, and hence, the $D^{3}$ should
perform fairly well even in frequency-selective fading channels. Nevertheless,
it can be noted that (\ref{E-DDD-00}) does not have a unique solution
because $\mathbf{d}$ and $-\mathbf{d}$ can minimize (\ref{E-DDD-00}).
To resolve the phase ambiguity problem, one or more pilot symbols
can be used as a part of the sequence $\mathbf{d}$\textbf{. }In such
scenarios, the performance of the $D^{3}$ will be affected indirectly
by the frequency selectivity of the channel because the capability
of the pilot to resolve the phase ambiguity depends on its fading
coefficient. Another advantage of using pilot symbols is that it will
not be necessary to detect the $N$ symbols simultaneously. Instead,
it will be sufficient to detect $\mathcal{K}$ symbols at a time,
which can be exploited to simplify the system design and analysis.

Using the same approach of the frequency domain, the $D^{3}$ can
be designed to work in the time domain as well by minimizing the channel
coefficients over two consecutive subcarriers, i.e., two subcarriers
with the same index over two consecutive OFDM symbols, which is also
applicable to single carrier systems. It can be also designed to work
in both time and frequency domains, where the detector can be described
as 
\begin{equation}
\mathbf{\hat{D}}_{\mathcal{L}\text{,}\mathcal{K}}\mathbf{=}\arg\min_{\mathbf{\tilde{\mathbf{D}}}_{\mathcal{L}\text{,}\mathcal{K}}}\text{ }J\left(\tilde{\mathbf{D}}_{\mathcal{L}\text{,}\mathcal{K}}\right)\label{eq:opt-D}
\end{equation}
where $\mathbf{D}_{\mathcal{L}\text{,}\mathcal{K}}$ is an $\mathcal{L}\times\mathcal{K}$
data matrix, $\mathcal{L}$ and $\mathcal{K}$ are the time and frequency
detection window size, and the objective function $J\left(\tilde{\mathbf{D}}\right)$
is given by 
\begin{equation}
J\left(\tilde{\mathbf{D}}_{\mathcal{L}\text{,}\mathcal{K}}\right)=\sum_{\ell=0}^{\mathcal{L}-1}\sum_{v=0}^{\mathcal{K}-2}\left\vert \frac{r_{v}^{\ell}}{\tilde{d}_{v}^{\ell}}-\frac{r_{\acute{v}}^{\ell}}{\tilde{d}_{\acute{v}}^{\ell}}\right\vert ^{2}+\left\vert \frac{r_{v}^{\ell}}{\tilde{d}_{v}^{\ell}}-\frac{r_{v}^{\acute{\ell}}}{\tilde{d}_{v}^{\acute{\ell}}}\right\vert ^{2}\text{.}\label{eq:objective-function}
\end{equation}
For example, if the detection window size is chosen to be the LTE
resource block, then, $\mathcal{L}=14$ and $\mathcal{K=}12$. Moreover,
the system presented in \eqref{eq:objective-function} can be extended
to the multi-branch receiver scenarios, single-input multiple-output
(SIMO) as, 
\begin{align}
\hat{\mathbf{D}} & =\arg\min_{\tilde{\mathbf{d}}}\sum_{n=1}^{\mathcal{N}}\sum_{\ell=0}^{\mathcal{L}-1}\sum_{v=0}^{\mathcal{K}-2}\left\vert \frac{r_{v}^{\ell,n}}{\tilde{d}_{v}}-\frac{r_{\acute{v}}^{\ell,n}}{\tilde{d}_{\acute{v}}^{\ell}}\right\vert ^{2}+\left\vert \frac{r_{v}^{\ell,n}}{\tilde{d}_{v}^{\ell}}-\frac{r_{v}^{\acute{\ell},n}}{\tilde{d}_{v}^{\acute{\ell}}}\right\vert ^{2}
\end{align}
where $\mathcal{N}$ is the number of receiving antennas.

\section{Efficient Implementation of $D^{3}$\label{sec:Efficient-Implementation-of-D3}}

\begin{figure}[t]
\begin{centering}
\includegraphics[scale=0.62]{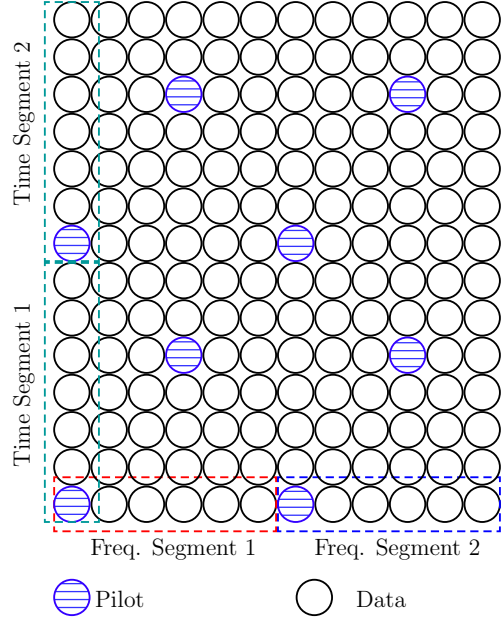} 
\par\end{centering}
\caption{Example of a 1-D segmentation over the frequency domain for an LTE-A
resource block.\label{fig:2D-to-1D} }
\vspace{-1.5em}
\end{figure}

It can be noted from \eqref{eq:opt-D} and \eqref{eq:objective-function}
that solving for $\hat{\mathbf{D}}$, given that $N_{P}$ pilot symbols
are used, requires an $M^{\mathcal{K}\mathcal{L-}N_{P}}$ trials if
brute force search is adopted, which is prohibitively complex, and
thus, reducing the computational complexity is crucial. \textcolor{black}{Towards
this goal, the two dimensional (2-D) resource block (RB) can be divided
into a number of one-dimensional (1-D) segments in time and frequency
domains in order to reduce the complexity from order $\mathcal{O}\left(M^{\mathcal{K}\times\mathcal{L-}N_{P}}\right)$
to $\mathcal{O}\left(M\mathcal{\times\left(\mathcal{\mathcal{K}L-}\mathit{N_{P}}\right)}\right)$.
In order words, the time complexity evolves exponentially as the detection
size increases in the 2-D block, while it grows linearly in the cascaded
1-D block, which is significant complexity reduction. Fig. \ref{fig:2D-to-1D}
shows an example of decomposing the 2-D LTE-A RB into several 1-D
segments over time and frequency. As can be noted from the figure,
the RB consists of $168$ subcarriers among which $8$ subcarriers
are pilots. It is worth noting that there are some rows and columns
in the RB that do not have pilots, and thus, the detection of the
entire block can be performed as described in Subsection \ref{subsec:Resource-Block-Detection}.}

\subsection{\label{subsec:The-Viterbi-Algorithm}The Viterbi Algorithm (VA)}

\begin{figure}

\begin{centering}
\includegraphics[scale=0.8]{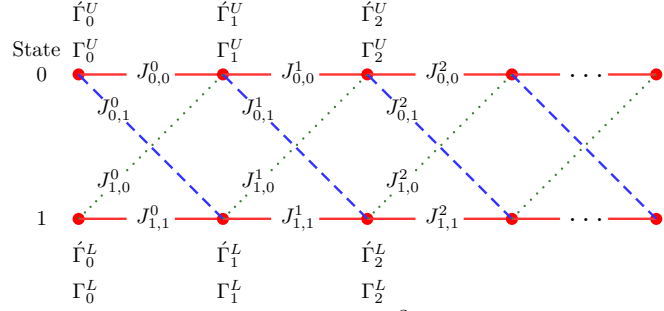}
\par\end{centering}
\caption{Trellis diagram of the $D^{3}$ detector for BPSK.\label{fig:Viterbi-D3}}
  \vspace{-1.5em} 
\end{figure}

By noting that the expression in (\ref{E-DDD-00}) corresponds to
the sum of correlated terms, which can be modeled as a first-order
Markov process, then MLSD techniques such as the VA can be used to
implement the $D^{3}$ efficiently. For example, the trellis diagram
of the VA with binary phase shift keying (BPSK) is shown in Fig. \ref{fig:Viterbi-D3},
and can be implemented as follows:
\begin{enumerate}
\item Initialize the path metrics $\left\{ \Gamma_{0}^{U},\acute{\Gamma}_{0}^{U},\Gamma_{0}^{L},\acute{\Gamma}_{0}^{L}\right\} =0$,
where $U$ and $L$ denote the upper and lower branches, respectively.
Since BPSK is used, the number of states is $2$.
\item Initialize the counter, $c=0$.
\item Compute the branch metric $J_{m,n}^{c}=\left\vert \frac{rc}{m}-\frac{r_{\acute{c}}}{n}\right\vert ^{2}$,
where $m$ is current symbol index, $m=0\rightarrow\tilde{d}=-1$,
and $m=1\rightarrow\tilde{d}=1$, and $n$ is the next symbol index
using the same mapping as $m$.
\item Compute the path metrics using the following rules,
\[
\begin{array}{ccc}
\Gamma_{\acute{c}}^{U}=\min\left[\Gamma_{c}^{U}\text{, }\acute{\Gamma}_{c}^{U}\right]+J_{00}^{c} &  & \Gamma_{\acute{c}}^{L}=\min\left[\Gamma_{c}^{L}\text{, }\acute{\Gamma}_{c}^{L}\right]+J_{01}^{c}\\
\acute{\Gamma}_{\acute{c}}^{U}=\min\left[\Gamma_{c}^{U}\text{, }\acute{\Gamma}_{c}^{U}\right]+J_{10}^{c} &  & \acute{\Gamma}_{\acute{c}}^{L}=\min\left[\Gamma_{c}^{L}\text{, }\acute{\Gamma}_{c}^{L}\right]+J_{11}^{c}
\end{array}
\]
\item Track the surviving paths, $2$ paths in the case of BPSK.
\item Increase the counter, $c=c+1$.
\item if $c=\mathcal{K}$, the algorithm ends. Otherwise, go to step 3.
\end{enumerate}

It is worth mentioning that placing a pilot symbol at the edge of
a segment terminates the trellis. To simplify the discussion, assume
that the pilot value is $-1$, and thus we compute only $J_{0,0}$
and $J_{1,0}$. Consequently, long data sequences can be divided into
smaller segments bounded by pilots, which can reduce the delay by
performing the detection over the sub-segments in parallel without
sacrificing the error rate performance.

\subsection{\textmd{\textcolor{black}{\normalsize{}Resource Block Detection\label{subsec:Resource-Block-Detection}}}}

\textcolor{black}{As can be noted from Fig. \ref{fig:2D-to-1D}, the
segmentation process can be applied directly to any row or column
given that has one or more pilots. Nevertheless, there are some rows
and columns that do not have pilots. In such scenarios, the detection,
for example, can be performed in two steps as follows:}
\begin{enumerate}
\item \textcolor{black}{Detect all rows (frequency domain subcarriers) with
pilots, i.e., rows 1, 5, 8 and 12.}
\item \textcolor{black}{As a result of the first step, each column (time
domain subcarrier) has either pilots, data symbols whose values are
known as a result of the detection in the first step, or both, as
in the case of columns 1, 4, 7 and 10. Therefore, all remaining subcarriers
can be detected using the symbols detected in the first step.}
\end{enumerate}
\textcolor{black}{It is worth noting that the number and distribution
of the pilot symbols in the RB impact the error rate performance,
power and spectral efficiency of the system. For example, the first
frequency segment shown in Fig. \ref{fig:2D-to-1D} consists of seven
subcarriers, two of them are allocated for pilots. By defining the
throughput, or the spectral efficiency, as the ratio of the number
of information symbols to the total number of symbols per segment,
then the throughput of the first frequency and time segments in Fig.
\ref{fig:2D-to-1D} is about 83.3\% and 85.7\%, respectively. Nevertheless,
the system throughput is determined by the total number of pilots
and information subcarriers within an RB rather than a segment. By
noting that there are only eight pilots among the 168 resource elements
, then the throughput loss is about $4.7\%$ and the throughput is
about 95.2\%. The same argument can be applied to the power efficiency
of the system where 4.7\% of the power will be allocated to pilots.}

\subsection{System Design with an Error Control Coding}

Forward error correction (FEC) coding can be integrated with the $D^{3}$
in two ways, based on the decoding process, i.e., hard or soft decision
decoding. For the hard decision decoding, the integration of FEC coding
is straightforward where the output of the $D^{3}$ is applied directly
to the hard decision decoder (HDD).

For the soft decision decoding, we can exploit the coded data to enhance
the performance of the $D^{3}$, and then use the $D^{3}$ output
to estimate the channel coefficients in a decision-directed manner.
The $D^{3}$ with coded data can be expressed as
\begin{equation}
\mathbf{\hat{d}}=\arg\min_{\tilde{\mathbf{u}}\in\mathbb{U}}\sum_{v=0}^{N-2}\left\vert \frac{r_{v}}{\tilde{u}_{v}}-\frac{r_{\acute{v}}}{\tilde{u}_{\acute{v}}}\right\vert ^{2}\label{E-D3-Joint}
\end{equation}
where $\mathbb{U}$ is the set of all codewords modulated using the
same modulation used at the transmitter. Therefore, the trial sequences
$\tilde{\mathbf{u}}$ are restricted to particular sequences. For
the case of convolutional codes, the detection and decoding processes
can be integrated smoothly since both of them are using the VA. Such
an approach can be adopted with linear block codes as well because
trellis-based decoding can be also applied to block codes \cite{Trellis-Block}.

\section{Error Rate Analysis of the $D^{3}$\label{sec:System-Performance-Analysis}}

The system BER analysis is presented for several cases according to
the pilot and data arrangements. For simplicity, each case is discussed
in separate subsections. To make the analysis tractable, we consider
BPSK modulation in the analysis while the BER of higher-order modulations
is obtained via Monte Carlo simulations.

\begin{figure}
\begin{minipage}[t]{0.45\columnwidth}%
\begin{center}
\includegraphics[width=0.48\columnwidth]{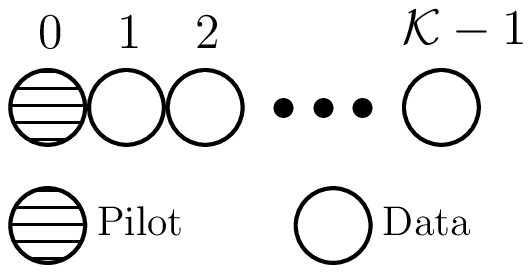}
\par\end{center}
\begin{center}
\caption{Single-sided pilot segment. \label{fig:Single-sided-pilot}}
  \vspace{-1.5em} 
\par\end{center}%
\end{minipage}\hfill{}%
\begin{minipage}[t]{0.45\columnwidth}%
\begin{center}
\includegraphics[width=0.48\columnwidth]{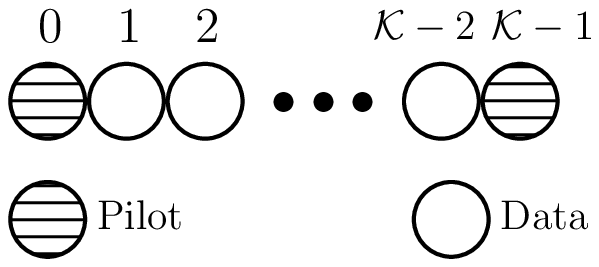}
\par\end{center}
\begin{center}
\caption{Double-sided pilot segment. \label{fig:Double-sided-pilot}}
  \vspace{-1.5em} 
\par\end{center}%
\end{minipage}
\end{figure}

\subsection{Single-Sided Pilot \label{subsec:Single-Sided-Pilot}}

To detect a data segment that contains $\mathcal{K}$ symbols, at
least one pilot symbol should be part of the segment in order to resolve
the phase ambiguity problem. Consequently, the analysis in this subsection
considers the case where there is only one pilot within the $\mathcal{K}$
symbols, as shown in Fig. \ref{fig:Single-sided-pilot}. Given that
the FFT output vector $\mathbf{r}=\left[r_{0}\text{, }r_{1}\text{,}\ldots,r_{N-1}\right]$
is divided into $L$ segments each of which consists of $\mathcal{K}$
symbols, including the pilot symbol, then the frequency domain $D^{3}$
detector can be written as, 
\begin{equation}
\hat{\mathbf{d}}_{l}=\arg\min_{\tilde{\mathbf{d}}}\sum_{v=l}^{\mathcal{K-}2+l}\left\vert \frac{r_{v}}{\tilde{d_{v}}}-\frac{r_{\acute{v}}}{\tilde{d}_{\acute{v}}}\right\vert ^{2}\,\,\,\,\mathcal{K}\in\left\{ 2,3,\dots,N-1\right\} \label{eq:d_hat}
\end{equation}
where $l$ denotes the index of the first subcarrier in the segment,
and without loss of generality, we consider that $l=0$. Therefore,
by expanding \eqref{eq:d_hat} we obtain, 
\begin{multline}
\hat{\mathbf{d}}_{0}=\arg\min_{\tilde{\mathbf{d}}}\left(\frac{r_{0}}{\tilde{d_{0}}}-\frac{r_{1}}{\tilde{d}_{1}}\right)\left(\frac{r_{0}}{\tilde{d_{0}}}-\frac{r_{1}}{\tilde{d}_{1}}\right)^{\ast}+\cdots+\left(\frac{r_{\mathcal{K}-2}}{\tilde{d}_{\mathcal{K}-2}}-\frac{r_{\mathcal{K}-1}}{\tilde{d}_{\mathcal{K}-1}}\right)\left(\frac{r_{\mathcal{K}-2}}{\tilde{d}_{\mathcal{K}-2}}-\frac{r_{\mathcal{K}-1}}{\tilde{d}_{\mathcal{K}-1}}\right)^{\ast}\label{eq:analysis-expansion-01}
\end{multline}
which can be simplified to, 
\begin{multline}
\hat{\mathbf{d}}_{0}=\arg\min_{\tilde{\mathbf{d}}}\left\vert \frac{r_{0}}{\tilde{d_{0}}}\right\vert ^{2}+\left\vert \frac{r_{1}}{\tilde{d_{1}}}\right\vert ^{2}+\dots+\left\vert \frac{r_{\mathcal{K}-1}}{\tilde{d}_{\mathcal{K}-1}}\right\vert ^{2}-\frac{r_{0}}{\tilde{d_{0}}}\frac{r_{1}}{\tilde{d}_{1}^{\ast}}-\frac{r_{0}}{\tilde{d_{0}^{\ast}}}\frac{r_{1}}{\tilde{d}_{1}}-\cdots\\
-\frac{r_{\mathcal{K}-2}}{\tilde{d}_{\mathcal{K}-2}}\frac{r_{\mathcal{K}-1}}{\tilde{d}_{\mathcal{K}-1}^{\ast}}-\frac{r_{\mathcal{K}-2}}{\tilde{d}_{\mathcal{K}-2}^{\ast}}\frac{r_{\mathcal{K}-1}}{\tilde{d}_{\mathcal{K}-1}}.\label{eq:analysis-expansion-02}
\end{multline}
For BPSK, $\left\vert r_{v}/\tilde{d_{v}}\right\vert ^{2}=\left\vert r_{v}\right\vert ^{2}$,
which is a constant term with respect to the maximization process
in \eqref{eq:analysis-expansion-02}, and thus, they can be dropped.
Therefore, the detector is reduced to 
\begin{equation}
\hat{\mathbf{d}}_{0}=\arg\max_{\tilde{\mathbf{d}_{0}}}\sum_{v=0}^{\mathcal{K-}2}\Re\left\{ \frac{r_{v}r_{\acute{v}}}{\tilde{d_{v}}\tilde{d}_{\acute{v}}}\right\} .
\end{equation}
Given that the pilot symbol is placed in the first subcarrier and
noting that $d_{v}\in\left\{ -1,1\right\} $, then $\tilde{d_{0}}=1$
and $\hat{\mathbf{d}}_{0}$ can be written as 
\begin{equation}
\hat{\mathbf{d}}_{0}=\arg\max_{\tilde{d}_{0}\notin\tilde{\mathbf{d}}_{0}}\frac{1}{\tilde{d_{1}}}\Re\left\{ r_{0}r_{1}\right\} +\sum_{v=1}^{\mathcal{K-}2}\frac{1}{\tilde{d_{v}}\tilde{d}_{\acute{v}}}\Re\left\{ r_{v}r_{\acute{v}}\right\} .\label{eq:d_hat_single_sided}
\end{equation}
The sequence error probability ($P_{S}$), conditioned on the channel
frequency response over the $\mathcal{K}$ symbols ($\mathbf{H}_{0})$
and the transmitted data sequence $\mathbf{d}_{0}$ can be defined
as, 
\begin{equation}
P_{S}|_{\mathbf{H}_{0},\mathbf{d}_{0}}\triangleq\left.\Pr\left(\hat{\mathbf{d}_{0}}\neq\mathbf{d}_{0}\right)\right\vert _{\mathbf{H}_{0},\mathbf{d}_{0}}\label{eq:SEP-definition}
\end{equation}
which can be also written in terms of the conditional probability
of correct detection $P_{C}$ as, 
\begin{equation}
P_{C}|_{\mathbf{H}_{0},\mathbf{d}_{0}}=1-\Pr\left(\hat{\mathbf{d}_{0}}=\mathbf{d}_{0}\right)\mid_{\mathbf{H}_{0},\mathbf{d}_{0}}.\label{eq:SEP-Analysis-01}
\end{equation}
Without loss of generality, we assume that $\mathbf{d}_{0}\mathbf{=}[1$,
$1$,\ldots$,1]\triangleq\mathbf{1}${} . Therefore, 
\begin{equation}
P_{C}|_{\mathbf{H}_{0},\mathbf{\mathbf{1}}}=\Pr\left(\sum_{v=0}^{\mathcal{K-}2}\Re\left\{ r_{v}r_{\acute{v}}\right\} =\max_{\tilde{\mathbf{d}_{0}}}\left\{ \sum_{v=0}^{\mathcal{K-}2}\frac{\Re\left\{ r_{v}r_{\acute{v}}\right\} }{\tilde{d_{v}}\tilde{d}_{\acute{v}}}\right\} \right).\label{eq:probability-correct-sequence}
\end{equation}
Since $\mathbf{d}_{0}$ has $\mathcal{K-}1$ data symbols, then there
are $2^{\mathcal{K-}1}$ trial sequences, $\tilde{\mathbf{d}}_{0}^{(0)}$,
$\tilde{\mathbf{d}}_{0}^{(1)}$,$\ldots$, $\tilde{\mathbf{d}}_{0}^{(\psi)}$,
where $\psi=2^{\mathcal{K-}1}-1$, and $\tilde{\mathbf{d}}_{0}^{(\psi)}\mathbf{=}[1$,
$1$,\ldots$,1]${} . The first symbol in every sequence is set to
$1$, which is the pilot symbol. By defining $\sum_{v=0}^{\mathcal{K-}2}\frac{\Re\left\{ r_{v}r_{\acute{v}}\right\} }{\tilde{d_{v}}\tilde{d}_{\acute{v}}}\triangleq A_{n}$,
where $\tilde{d_{v}}\tilde{d}_{\acute{v}}\in\tilde{\mathbf{d}}_{0}^{(n)}$,
then (\ref{eq:probability-correct-sequence}) can be written as, 
\begin{equation}
P_{C}|_{\mathbf{H}_{0},\mathbf{\mathbf{1}}}=\Pr\left(A_{\psi}>A_{\psi-1},A_{\psi-2},\ldots,A_{0}\right)\label{E-PC-00}
\end{equation}
which, as depicted in Appendix I, can be simplified to 
\begin{equation}
P_{C}|_{\mathbf{H}_{0},\mathbf{\mathbf{1}}}=\prod\limits _{v=0}^{\mathcal{K-}2}\Pr\left(\Re\left\{ r_{v}r_{\acute{v}}\right\} >0\right).\label{eq:pc-expansion-2}
\end{equation}

To evaluate $P_{C}|_{\mathbf{H}_{0},\mathbf{\mathbf{1}}}$ given in
\eqref{eq:pc-expansion-2}, it is necessary to compute $\Pr\left(\Re\left\{ r_{v}r_{\acute{v}}\right\} >0\right)$,
which can be written as 
\begin{equation}
\Pr\left(\Re\left\{ r_{v}r_{\acute{v}}\right\} >0\right)=\Pr\left(\underbrace{r_{v}^{I}r_{\acute{v}}^{I}-r_{v}^{Q}r_{\acute{v}}^{Q}}_{r_{v,\acute{v}}^{\mathrm{SP}}}>0\right).\label{E-rSP}
\end{equation}
Given that $\mathbf{d}_{0}\mathbf{=}[1$, $1$,\ldots$,1]${} , then
$r_{v}^{I}=\Re\left\{ r_{v}\right\} =H_{v}^{I}+w_{v}^{I}$ and $r_{v}^{Q}=\Im\left\{ r_{v}\right\} =H_{v}^{Q}+w_{v}^{Q}$.
Therefore, $r_{v}^{I},$ $r_{v}^{Q}$, $r_{\acute{v}}^{I}$ and $r_{\acute{v}}^{Q}$
are independent conditionally Gaussian random variables with averages
$H_{v}^{I}$, $H_{v}^{Q}$, $H_{\acute{v}}^{I}$ and $H_{\acute{v}}^{Q}$,
respectively, and the variance for all elements is $\sigma_{w}^{2}$.
To derive the PDF of $r_{v,\acute{v}}^{\mathrm{SP}}$, the PDFs of
$r_{v}^{I}r_{\acute{v}}^{I}$ and $r_{v}^{Q}r_{\acute{v}}^{Q}$ should
be evaluated, where each of which corresponds to the product of two
Gaussian random variables. Although the product of two Gaussian variables
is not usually Gaussian, the limit of the moment-generating function
of the product has Gaussian distribution. Therefore, the product of
two variables $X\sim\mathcal{N}(\mu_{x},\sigma_{x}^{2})$ and $Y\sim\mathcal{N}(\mu_{y},\sigma_{y}^{2})$
tends to be $\mathcal{N}(\mu_{x}\mu_{y},\mu_{x}^{2}\sigma_{y}^{2}+\mu_{y}^{2}\sigma_{x}^{2})$
as the ratios $\mu_{x}/\sigma_{x}$ and $\mu_{y}/\sigma_{y}$ increase
\cite{Product of 2RV}. By noting that in in (\ref{E-rSP}) $\mathrm{E}\left[r_{y}^{x}\right]=H_{y}^{x}$,
$x\in\left\{ I,Q\right\} $ and $y\in\left\{ v,\acute{v}\right\} $
and $\sigma_{r_{y}^{x}}=\sigma_{w}$, thus $\mathrm{E}\left[r_{y}^{x}\right]/\sigma_{r_{y}^{x}}\gg1$
$\forall\left\{ x,y\right\} $. Moreover, because the PDF of the sum
or difference of two Gaussian random variables is also Gaussian, then,
$r_{v,\acute{v}}^{\mathrm{SP}}\sim\mathcal{N}\left(\bar{\mu}_{\mathrm{SP}},\bar{\sigma}_{\mathrm{SP}}^{2}\right)$
where $\bar{\mu}_{\mathrm{SP}}=H_{v}^{I}H_{\acute{v}}^{I}+H_{v}^{Q}H_{\acute{v}}^{Q}$
and $\bar{\sigma}_{\mathrm{SP}}^{2}=\sigma_{w}^{2}\left(\left\vert H_{v}\right\vert ^{2}+\left\vert H_{\acute{v}}\right\vert ^{2}+\sigma_{w}^{2}\right)$.
Consequently, 
\begin{equation}
P_{C}|_{\mathbf{H}_{0},\mathbf{1}}=\prod_{v=0}^{\mathcal{K}-2}\Pr\left(r_{v,\acute{v}}^{\mathrm{SP}}>0\right)=\prod_{v=0}^{\mathcal{K}-2}\left[1-Q\left(\sqrt{\frac{2\bar{\mu}_{\mathrm{SP}}}{\bar{\sigma}_{\mathrm{SP}}^{2}}}\right)\right]
\end{equation}
and 
\begin{equation}
P_{S}|_{\mathbf{H}_{0},\mathbf{1}}=1-\prod_{v=0}^{\mathcal{K}-2}\left[1-Q\left(\sqrt{\frac{2\bar{\mu}_{\mathrm{SP}}}{\bar{\sigma}_{\mathrm{SP}}^{2}}}\right)\right]\label{eq:SEP}
\end{equation}
where $Q\left(x\right)\triangleq\frac{1}{\sqrt{2\pi}}\int_{x}^{\infty}\exp\left(-\frac{t^{2}}{2}\right)dt$.
Since $H_{v}^{I}$ and $H_{v}^{Q}$ are independent, then, the condition
on $\mathbf{H}_{0}$ in \eqref{eq:SEP} can be removed by averaging
$P_{S}$ over the PDF of $\mathbf{H}_{0}^{I}$ and $\mathbf{H}_{0}^{Q}$
as, 
\begin{multline}
\mathrm{SEP}\mid_{\mathbf{d}=1}=\underbrace{\int_{-\infty}^{\infty}\int_{-\infty}^{\infty}\dots\int_{-\infty}^{\infty}}_{2\mathcal{K}\text{ fold}}\mathrm{SEP}\mid_{\mathbf{H}_{0},\mathbf{d}=1}f_{\mathbf{H}_{0}^{I}}\left(H_{0}^{I},H_{1}^{I},\dots,H_{\mathcal{K}-1}^{I}\right)\times\\
f_{\mathbf{H}_{0}^{Q}}\left(H_{0}^{Q},H_{1}^{Q},\dots,H_{\mathcal{K}-1}^{Q}\right)dH_{0}^{I}dH_{1}^{I}\dots dH_{\mathcal{K}-1}^{I}dH_{0}^{Q}dH_{1}^{Q}\dots dH_{\mathcal{K}-1}^{Q}\text{.}\label{eq:unconditional-SER}
\end{multline}
Because the random variables $H_{i}^{I}$ and $H_{i}^{Q}$ $\forall i$
in \eqref{eq:unconditional-SER} are real and Gaussian, their PDFs
are multivariate Gaussian distributions \cite{Proakis-Book-2001},
\begin{equation}
f_{\mathbf{X}}\left(X_{0},X_{1},\dots,X_{\mathcal{K}-1}\right)=\frac{\exp\left(-\frac{1}{2}(\mathbf{X}-\boldsymbol{\mu})^{\mathrm{T}}\boldsymbol{\Sigma}^{-1}(\mathbf{X}-\boldsymbol{\mu})\right)}{\sqrt{(2\pi)^{\mathcal{K}}|\boldsymbol{\Sigma}|}}\label{eq:multi-variate-gaussian}
\end{equation}
where $\boldsymbol{\mu}$ is the mean vector, which is defined as,
\begin{equation}
\boldsymbol{\mu}=\mathrm{E}\left[\mathbf{X}\right]=\left[\mathrm{E}\left[X_{1}\right],\mathrm{E}\left[X_{2}\right],\dots,\mathrm{E}\left[X_{\mathcal{K}-1}\right]\right]^{T}
\end{equation}
and $\boldsymbol{\Sigma}$ is the covariance matrix, $\boldsymbol{\Sigma}=\mathrm{E}\left[\left(\mathbf{X}-\mu\right)\left(\mathbf{X}-\mu\right)^{T}\right].$

Due to the difficulty of evaluating $2\mathcal{K}$ integrals, we
consider the special case of flat fading, which implies that $H_{v}=H_{\acute{v}}\triangleq H$
and $\left(H^{I}\right)^{2}+\left(H^{Q}\right)^{2}\triangleq\alpha^{2}$,
where $\alpha$ is the channel fading envelope, $\alpha=\left\vert H\right\vert $.
Therefore, the SEP expression in \eqref{eq:SEP} becomes, 
\begin{equation}
P_{S}|_{\alpha,\mathbf{1}}=1-\left[1-Q\left(\sqrt{\frac{\alpha^{2}}{\sigma_{w}^{2}\left(\alpha^{2}+\sigma_{w}^{2}\right)}}\right)\right]^{\mathcal{K}-1}.\label{eq:SEP-conditional-general}
\end{equation}
Recalling the Binomial Theorem, we get 
\begin{equation}
\left(a+b\right)^{n}=\sum_{v=0}^{n}\binom{n}{v}a^{n-v}b^{v}\text{, }\binom{n}{v}\triangleq\frac{n!}{\left(n-v\right)!v!}.\label{eq:binomial-theorem}
\end{equation}
Then the SEP formula in \eqref{eq:SEP-conditional-general} using
the Binomial Theorem in \eqref{eq:binomial-theorem} can be written
as, 
\begin{equation}
P_{S}|_{\alpha,\mathbf{1}}=1-\sum_{v=0}^{\mathcal{K}-1}\binom{\mathcal{K}-1}{v}\left(-1\right)^{v}\left[Q\left(\sqrt{\frac{\alpha^{2}}{\sigma_{w}^{2}\left(\alpha^{2}+\sigma_{w}^{2}\right)}}\right)\right]^{v}.\label{eq:sep_cond_higher_k}
\end{equation}
The conditioning on $\alpha$ can be removed by averaging over the
PDF of $\alpha$, which is Rayleigh. Therefore, 
\begin{equation}
f\left(\alpha\right)=\frac{\alpha}{\sigma_{H}^{2}}e^{-\frac{\alpha^{2}}{2\sigma_{H}^{2}}}.\label{eq:rayleigh-pdf}
\end{equation}
And hence, 
\begin{equation}
P_{S}|_{\mathbf{1}}=\int_{0}^{\infty}P_{S}|_{\alpha,\mathbf{1}}\text{ }f\left(\alpha\right)d\alpha.\label{E-Averaging}
\end{equation}
Because the expression in \eqref{eq:SEP-conditional-general} contains
high order of $Q$-function $Q^{n}\left(x\right)$, evaluating the
integral analytically becomes intractable for $\mathcal{K}>2$. For
the special case of $\mathcal{K}=2$, $P_{S}$ can be evaluated by
substituting (\ref{eq:sep_cond_higher_k}) and (\ref{eq:rayleigh-pdf})
into (\ref{E-Averaging}) and evaluating the integral yields the following
simple expression, 
\begin{equation}
P_{S}|_{\mathbf{1}}=\frac{1}{2\left(\bar{\gamma}_{s}+1\right)}\text{, \ \ }\bar{\gamma}_{s}\triangleq\frac{\mathrm{E}\left[\left\vert d_{v}\right\vert ^{2}\right]\mathrm{E}\left[\left\vert H\right\vert ^{2}\right]}{2\sigma_{w}^{2}}\label{E-Pe_K2}
\end{equation}
where $\bar{\gamma}_{s}$ is the average signal-to-noise ratio (SNR).
Moreover, because all data sequences have an equal probability of
error, then $P_{S}|_{\mathbf{1}}=P_{S}$, which also equivalent to
the bit error rate (BER). It is interesting to note that (\ref{E-Pe_K2})
is similar to the BER of the differential binary phase shift keying
(DBPSK) \cite{Proakis-Book-2001}. However, the two techniques are
essentially different as $D^{3}$ does not require differential encoding,
has no constraints on the shape of the signal constellation, and performs
well even in frequency-selective fading channels.

To evaluate $P_{S}$ for $\mathcal{K}>2$, we use an approximation
for $Q\left(x\right)$ in \cite{Q-Func-Approx-02}, which is given
by 
\begin{equation}
Q\left(x\right)\approx\frac{1}{\sqrt{2\pi\left(x^{2}+1\right)}}e^{-\frac{1}{2}x^{2}},\text{ }x\in\lbrack0,\infty).\label{eq:Q-func-Approx}
\end{equation}
Therefore, by substituting \eqref{eq:Q-func-Approx} into the conditional
SEP \eqref{eq:sep_cond_higher_k} and averaging over the Rayleigh
PDF \eqref{eq:rayleigh-pdf}, the evaluation of the SEP becomes straightforward.
For example, evaluating the integral for $\mathcal{K}=3$ gives, 
\begin{equation}
P_{S}|_{\mathbf{1}}=\frac{\zeta_{1}}{\pi}\mathit{\mathrm{Ei}}\left(1,\zeta_{1}+1\right)e^{\zeta_{1}+1}\text{, \ \ }\zeta_{1}\triangleq\frac{1}{2\bar{\gamma}_{s}}\left(\frac{1}{\bar{\gamma}_{s}}+1\right)
\end{equation}
where $\mathrm{Ei}\left(x\right)$ is the exponential integral (EI),
$\mathrm{Ei}\left(x\right)\triangleq-\int_{-x}^{\infty}\frac{e^{-t}}{t}dt$.
Similarly, $P_{S}$ for $\mathcal{K}=7$ can be evaluated to, 
\begin{equation}
P_{S}|_{\mathbf{1}}=\frac{\zeta_{2}}{64\pi^{3}}\left[e^{\zeta+3}\left(2\zeta_{2}+6\right)^{2}\text{ }\mathit{\mathrm{Ei}}\left(1,\zeta_{2}+3\right)-4\left(\zeta_{2}+1\right)\right]\text{, \ }\zeta_{2}\triangleq\frac{1}{2\bar{\gamma}_{s}}\left(\frac{1}{4\bar{\gamma}_{s}}+1\right).
\end{equation}

Although the SEP is a very useful indicator for the system error probability
performance, the BER is actually more informative. For a sequence
that contains $\mathcal{K}_{D}$ information bits, the BER can be
expressed as $P_{B}=\frac{1}{\Lambda}P_{S}$, where $\Lambda$ denotes
the average number of bit errors given a sequence error, which can
be defined as 
\begin{equation}
\Lambda=\sum_{m=1}^{\mathcal{K}_{D}}m\Pr\left(m\right).
\end{equation}
Because the SEP is independent of the transmitted data sequence, then,
without loss of generality, we assume that the transmitted data sequence
is $\mathbf{d}_{0}^{(0)}$. Therefore, 
\begin{equation}
\Lambda=\sum_{m=1}^{\mathcal{K}_{D}}m\Pr\left(\left\Vert \mathbf{\hat{d}}_{0}\right\Vert ^{2}=m\right)
\end{equation}
where $\left\Vert \mathbf{\hat{d}}_{0}\right\Vert ^{2}$, in this
case, corresponds to the Hamming weight of the detected sequence $\mathbf{\hat{d}}_{0}$,
which can be expressed as 
\begin{equation}
\Pr\left(\left\Vert \mathbf{\hat{d}}_{0}\right\Vert ^{2}=m\right)=\Pr\left(\mathbf{d}_{0}^{(0)}\rightarrow\bigcup\limits _{i}\mathbf{d}_{0}^{(i)}\right)\text{, }\left\Vert \mathbf{d}_{0}^{(i)}\right\Vert ^{2}=m
\end{equation}
where $\mathbf{d}_{0}^{(0)}\rightarrow\mathbf{d}_{0}^{(i)}$ denotes
the pairwise error probability (PEP). By noting that $\Pr\left(\mathbf{d}_{0}^{(0)}\rightarrow\mathbf{d}_{0}^{(i)}\right)\neq\Pr\left(\mathbf{d}_{0}^{(0)}\rightarrow\mathbf{d}_{0}^{(j)}\right)$
$\forall i\neq j$, then deriving the PEP for all cases of interest
is intractable. As an alternative, a simple approximation is derived.

For a sequence that consists of $\mathcal{K}_{D}$ information bits,
the BER is bounded by 
\begin{equation}
\frac{1}{\mathcal{K}_{D}}P_{S}\leq P_{B}\leq P_{S}\text{.}\label{E-Bounds}
\end{equation}
In practical systems, the number of bits in the detected sequence
is generally not large, which implies that the upper and lower bounds
in (\ref{E-Bounds}) are relatively tight, and hence, the BER can
be approximated as the middle point between the two bounds as, 
\begin{equation}
P_{B}\approx\frac{P_{S}}{0.5\left(1+\mathcal{K}_{D}\right)}.\label{E_PB}
\end{equation}

The analysis of the general $1\times\mathcal{N}$ SIMO system is a
straightforward extension of the single-input single-output (SISO)
case. To simplify the analysis, we consider the flat channel case
where the conditional SEP can be written as, 
\begin{equation}
P_{S}|_{\mathbf{\alpha}}=1-\left[1-Q\left(\sqrt{\frac{\sum_{i=1}^{\mathcal{N}}\alpha_{i}^{2}}{\sigma_{w}^{2}\left(\mathcal{N}\sigma_{w}^{2}+\sum_{i=1}^{\mathcal{N}}\alpha_{i}^{2}\right)}}\right)\right]^{\mathcal{K}-1}.
\end{equation}
Given that all the receiving branches are independent, the fading
envelopes will have Rayleigh distribution $\alpha_{i}\sim\mathcal{R}\left(2\sigma_{H}^{2}\right)$
$\forall i$, and thus, $\sum_{i=1}^{\mathcal{N}}\alpha_{i}^{2}\triangleq a$
will have Gamma distribution, $a\sim\mathcal{G}\left(\mathcal{N},2\sigma_{H}^{2}\right)$,
\begin{equation}
f\left(a\right)=\left(2\sigma_{H}^{2}\right)^{\mathcal{N}}e^{-2\sigma_{H}^{2}a}\frac{_{a^{\mathcal{N}-1}}}{\Gamma\left(\mathcal{N}\right)}.
\end{equation}
Therefore, the unconditional SEP can be evaluated as, 
\begin{equation}
P_{S}=\int_{0}^{\infty}P_{S}|_{\mathbf{\alpha}}\text{ }f_{A}\left(a\right)da.
\end{equation}
For the special case of $\mathcal{N=}2$, $\mathcal{K}=2$, $P_{S}$
can be evaluated as, 
\begin{equation}
P_{S}=\frac{1}{2}+Q\left(\frac{\varkappa}{\sqrt{\bar{\gamma}_{s}}}\right)\left[2\bar{\gamma}_{s}\left(\frac{\bar{\gamma}_{s}}{\sqrt{2}}+2\right)-e^{\varkappa^{2}}\right]-\bar{\gamma}_{s}\frac{\varkappa}{\sqrt{2\pi}}
\end{equation}
where $\varkappa\triangleq\sqrt{2+\bar{\gamma}_{s}}.$ Computing the
closed-form formulas for other values of $\mathcal{N}$ \ and $\mathcal{K}$
can be evaluated following the same approach used in the SISO case.

\subsection{Double-Sided Pilot \label{subsec:Double-Sided-Pilot}}

Embedding more pilots in the detection segment can improve the detector's
performance. Consequently, it worth investigating the effect of embedding
more pilots in the SEP analysis. More specifically, we consider double-sided
segment, $\tilde{d}_{0}=1$, $\tilde{d}_{\mathcal{K}-1}=1$, as illustrated
in Fig. \ref{fig:Double-sided-pilot}. In this case, the detector
can be expressed as, 
\begin{equation}
\hat{\mathbf{d}_{0}}=\arg\max_{\tilde{\mathbf{d}}_{0}}\frac{\Re\left\{ r_{0}r_{1}\right\} }{\tilde{d}_{1}}+\frac{\Re\left\{ r_{\mathcal{K}-2}r_{\mathcal{K}-1}\right\} }{\tilde{d}_{\mathcal{K}-2}}+\sum_{v=1}^{\mathcal{K-}3}\frac{\Re\left\{ r_{v}r_{\acute{v}}\right\} }{\tilde{d_{v}}\tilde{d}_{\acute{v}}},\text{\thinspace\thinspace\thinspace\thinspace\ensuremath{\mathcal{K}\in\left\{ 3,4,\dots,N-1\right\} .}}\label{eq:d_hat_double_sided}
\end{equation}
From the definition in \eqref{eq:d_hat_double_sided}, the probability
of receiving the correct sequence can be derived based on the reduced
number of trials as compared to \eqref{eq:d_hat_single_sided}. Therefore,
\begin{multline}
P_{C}|_{\mathbf{H}_{0},\mathbf{1}}=\Pr\Big(\left(\Re\left\{ r_{0}r_{1}\right\} +\Re\left\{ r_{\mathcal{K}-2}r_{\mathcal{K}-1}\right\} \right)\cap\\
\Re\left\{ r_{1}r_{2}\right\} \cap\Re\left\{ r_{2}r_{3}\right\} \cap\dots\cap\Re\left\{ r_{\mathcal{K}-4}r_{\mathcal{K}-3}\right\} >0\Big)\label{eq:pc-double-sided}
\end{multline}
which, similar to the single-sided case, can be written as, 
\begin{equation}
P_{C}|_{\mathbf{H}_{0},\mathbf{1}}=\Pr\left(\left[\prod_{v=0}^{\mathcal{K}-3}\Pr\left(\Re\left\{ r_{v}r_{\acute{v}}\right\} \right)+\prod_{v=1}^{\mathcal{K}-2}\Pr\left(\Re\left\{ r_{v}r_{\acute{v}}\right\} \right)\right]>0\right).
\end{equation}
Therefore, 
\begin{equation}
P_{S}|_{\mathbf{H}_{0},\mathbf{1}}=1-\left[1-Q\left(\sqrt{\frac{2\sqrt{2}\bar{\mu}_{\mathrm{SP}}}{\bar{\sigma}_{\mathrm{SP}}^{2}}}\right)\right]\times\prod_{v=1}^{\mathcal{K}-3}\left[1-Q\left(\sqrt{\frac{2\bar{\mu}_{\mathrm{SP}}}{\bar{\sigma}_{\mathrm{SP}}^{2}}}\right)\right].\label{eq:SEP-1}
\end{equation}
For flat fading channels, the SEP expression in \eqref{eq:SEP-1}
can be simplified by following the same procedure in Subsection \ref{subsec:Single-Sided-Pilot},
for the special case of $\mathcal{K}=3$, the SEP becomes, 
\begin{equation}
P_{S}=\left(\frac{\Upsilon}{2}-\sqrt{2}\right)\frac{1}{\Upsilon}\text{, \ }\Upsilon\triangleq\sqrt{8\bar{\gamma}_{s}+\sqrt{2}\left(4+\frac{1}{\bar{\gamma}_{s}}\right)}.
\end{equation}
For $\mathcal{K}>3$, the approximation of $Q^{n}\left(x\right)$,
as illustrated in Subsection \ref{subsec:Single-Sided-Pilot}, can
be used in \eqref{eq:SEP-1} to average over the PDF in \eqref{eq:rayleigh-pdf}.
For example, the case $\mathcal{K}=4$ can be evaluated as, 
\begin{equation}
P_{S}=\frac{1}{8\pi\bar{\gamma}_{s}}\left(\Omega_{1}-1\right)e^{\Omega_{1}}\mathit{\mathrm{Ei}}\left(1,\Omega_{1}\right)\text{, \ }\Omega_{1}\triangleq1+\frac{\sqrt{2}}{4\bar{\gamma}_{s}}\left(1+\frac{1}{4\bar{\gamma}_{s}}\right).
\end{equation}
For $\mathcal{K}=6$, 
\begin{equation}
P_{S}=\frac{\Omega_{1}-1}{4\pi^{2}}\left[1-\left[\left(\Omega_{1}-1\right)e^{\Omega_{2}}+2\right]\mathit{\mathrm{Ei}}\left(1,\Omega_{2}\right)\right]\text{, \ }\Omega_{2}\triangleq2+\frac{\sqrt{2}}{\bar{\gamma}_{s}}\left(8+\frac{1}{32\bar{\gamma}_{s}}\right)
\end{equation}
For the double-sided pilot, $P_{B}=P_{S}$ for the case of $\mathcal{K}=3$,
while it can be computed using (\ref{E_PB}) for $\mathcal{K}>3$.

\section{Complexity Analysis\label{sec:Complexity-Analysis}}

The computational complexity is evaluated as the total number of primitive
operations needed to perform the detection. The operations that will
be used are the number of real additions ($R_{A}$), real multiplications
($R_{M}$), and real divisions ($R_{D}$) required to produce the
set of detected symbols $\hat{\mathbf{d}}$ for each technique. It
worth noting that one complex multiplication ($C_{M}$) is equivalent
to four $R_{M}$ and three $R_{A}$ operations, while one complex
addition ($C_{A}$) requires two $R_{A}$. To simplify the analysis,
we first assume that constant modulus (CM) constellations such as
MPSK is used, then, we evaluate the complexity for higher-order modulation
such as quadrature amplitude modulation (QAM) modulation.

\subsection{Complexity of Conventional OFDM Detectors\label{subsec:Complexity-of-Conventional}}

The complexity of the conventional OFDM receiver that consists of
the following main steps with the corresponding computational complexities:
\begin{enumerate}
\item Channel estimation of the pilot symbols, which computes $\hat{H}_{k}$
at all pilot subcarriers. Assuming that the pilot symbol $d_{k}$
is selected from a CM constellation, then $\hat{H}_{k}=r_{k}d_{k}^{*}$
and hence, $N_{P}$ complex multiplications are required. Therefore,
$R_{A}^{\left(1\right)}=4N_{P}$ and $R_{M}^{\left(1\right)}=4N_{P}$.
\item Interpolation, which is used to estimate the channel at the non-pilot
subcarriers. The complexity of the interpolation process depends on
the interpolation algorithm used. For comparison purposes, we assume
that linear interpolation is used, which is the least complex interpolation
algorithm. The linear interpolation requires one complex multiplication
and two complex additions per interpolated sample. Therefore, the
number of complex multiplications required is $N-N_{P}$ and the number
of complex additions is $2\left(N-N_{P}\right)$. And hence, $R_{A}^{\left(2\right)}=7\left(N-N_{P}\right)$
and $R_{M}^{\left(2\right)}=4\left(N-N_{P}\right)$.
\item Equalization, a single-tap equalizer requires $N-N_{P}$ complex division
to compute the decision variables $\check{r}_{k}=\frac{r_{k}}{\hat{H}_{k}}=r_{k}\frac{\hat{H}_{k}^{*}}{\left|\hat{H}_{k}^{*}\right|^{2}}$.
Therefore, one complex division requires two complex multiplications
and one real division. Therefore, $R_{A}^{\left(3\right)}=6\left(N-N_{P}\right)$,
$R_{M}^{\left(3\right)}=8\left(N-N_{P}\right)$ and $R_{D}^{\left(3\right)}=\left(N-N_{P}\right)$.
\item Detection, assuming symbol-by-symbol minimum distance detection, the
detector can be expressed as $\hat{d}_{k}=\arg\min_{\tilde{d}_{i}}J\left(\tilde{d}_{i}\right),\,\,\forall i\in\left\{ 0,1,\dots,M-1\right\} $
where $J\left(\tilde{d}_{i}\right)=\left|\check{r}_{k}-\tilde{d}_{i}\right|^{2}$
. Assuming CM modulation is used, expanding the cost function and
dropping the constant terms we can write $J\left(\tilde{d}_{k}\right)=-\check{r}_{k}\tilde{d}_{k}^{*}-\check{r}_{k}^{*}\tilde{d}_{k}$.
We can also drop the minus sign from the cost function, and thus,
the objective becomes maximizing the cost function $\hat{d}_{k}=\arg\min_{\tilde{d}_{i}}J\left(\tilde{d}_{i}\right)$.
Since the two terms are complex conjugate pair, then $-\check{r}_{k}\tilde{d}_{k}^{*}-\check{r}_{k}^{*}\tilde{d}_{k}=2\Re\left\{ \check{r}_{k}\tilde{d}_{k}^{*}\right\} $,
and thus we can write the detected symbols as, 
\begin{equation}
\hat{d}_{k}=\arg\max_{\tilde{d}_{k}}\left(\Re\left\{ \check{r}_{k}\right\} \Re\left\{ \tilde{d}_{k}^{*}\right\} -\Im\left\{ \check{r}_{k}\right\} \Im\left\{ \tilde{d}_{k}^{*}\right\} \right)
\end{equation}
Therefore, the number of real multiplications required for each information
symbol is $2M$, and the number of additions is $M$. Therefore, $R_{A}^{\left(4\right)}=\left(N-N_{P}\right)M$
and $R_{M}^{\left(4\right)}=2\left(N-N_{P}\right)M$. 
\end{enumerate}
Finally, the total computational complexity per OFDM symbol can be
obtained by adding the complexities of the individual steps $1\rightarrow4$,
as:

\begin{align}
R_{A}^{CM} & ={\displaystyle \sum_{i=1}^{4}R_{A}^{\left(i\right)}=\left(13+M\right)N-\left(10+M\right)N_{P}}\\
R_{M}^{CM} & =\sum_{i=1}^{4}R_{M}^{\left(i\right)}=2N\left(6+M\right)-2N_{P}\left(4+M\right)\\
R_{D}^{CM} & =\sum_{i=1}^{4}R_{D}^{\left(i\right)}=N-N_{P}.
\end{align}

\textcolor{black}{For higher modulation orders, such as QAM, the complexity
of the conventional OFDM receivers considering addition division operations
is computed following the same steps $1\rightarrow4$ above, and found
to be as:}

\textcolor{black}{
\begin{align}
R_{A}^{QAM} & ={\displaystyle \sum_{i=1}^{4}R_{A}^{\left(i\right)}=}6N_{P}+\left(13+2M\right)\left(N-N_{P}\right)\\
R_{M}^{QAM} & =\sum_{i=1}^{4}R_{M}^{\left(i\right)}=8N_{P}+\left(12+4M\right)\left(N-N_{P}\right)\\
R_{D}^{QAM} & =\sum_{i=1}^{4}R_{D}^{\left(i\right)}=N_{P}+2M\left(N-N_{P}\right)
\end{align}
}

\subsection{Complexity of the $D^{3}$}

The complexity of the $D^{3}$ based on the VA is mostly determined
by the branch and path metrics calculation. The branch metrics can
be computed as 
\begin{equation}
J_{m,n}^{c}=\frac{\left\vert r_{c}\right\vert ^{2}}{\left\vert \tilde{d}_{m}\right\vert ^{2}}-\frac{r_{c}r_{\acute{c}}^{\ast}}{\tilde{d}_{m}\tilde{d}_{n}^{\ast}}-\frac{r_{c}^{\ast}r_{\acute{c}}}{\tilde{d}_{m}^{\ast}\tilde{d}_{n}}+\frac{\left\vert r_{c}\right\vert ^{2}}{\left\vert \tilde{d}_{n}\right\vert ^{2}}.
\end{equation}
For CM constellation, the first and last terms are constants, and
hence, can be dropped. Therefore, 
\begin{equation}
J_{m,n}^{c}=-\frac{r_{c}r_{\acute{c}}^{\ast}}{\tilde{d}_{m}\tilde{d}_{n}^{\ast}}+\frac{r_{c}^{\ast}r_{\acute{c}}}{\tilde{d}_{m}^{\ast}\tilde{d}_{n}}.\label{eq:branch-metric-viterbi}
\end{equation}
By noting that the two terms in \eqref{eq:branch-metric-viterbi}
are the complex conjugate pair, then 
\begin{equation}
J_{m,n}^{c}=-2\Re\left\{ \frac{r_{c}r_{\acute{c}}^{\ast}}{\tilde{d}_{m}\tilde{d}_{n}^{\ast}}\right\} .\label{eq:branch-metric-viterbi-02}
\end{equation}
From the expression in \eqref{eq:branch-metric-viterbi-02}, the constant
``$-2$\textquotedblright{} can be dropped from the cost function,
however, the problem with be flipped to a maximization problem. Therefore,
by expanding \eqref{eq:branch-metric-viterbi-02}, we get, 
\begin{equation}
J_{m,n}^{c}=\Re\left\{ \frac{\Re\left\{ r_{c}\right\} \Re\left\{ r_{\acute{c}}^{\ast}\right\} -\Im\left\{ r_{c}\right\} \Im\left\{ r_{\acute{c}}^{\ast}\right\} +j\left[-\Re\left\{ r_{c}\right\} \Im\left\{ r_{\acute{c}}^{\ast}\right\} +\Im\left\{ r_{c}\right\} \Im\left\{ r_{\acute{c}}^{\ast}\right\} \right]}{\Re\left\{ \tilde{d}_{m}\tilde{d}_{n}^{\ast}\right\} +j\Im\left\{ \tilde{d}_{m}\tilde{d}_{n}^{\ast}\right\} }\right\} .\label{eq:branch-metric-03}
\end{equation}
By defining $\tilde{d}_{m}\tilde{d}_{n}^{\ast}\triangleq\tilde{u}_{m,n},$
and using complex numbers identities, we get \eqref{eq:branch-metric-04},
\begin{equation}
J_{m,n}^{c}=\frac{\left[\Re\left\{ r_{c}\right\} \Re\left\{ r_{\acute{c}}^{\ast}\right\} +\Im\left\{ r_{c}\right\} \Im\left\{ r_{\acute{c}}^{\ast}\right\} \right]\Re\left\{ \tilde{u}_{m,n}\right\} -\left[-\Re\left\{ r_{c}\right\} \Im\left\{ r_{\acute{c}}^{\ast}\right\} +\Im\left\{ r_{c}\right\} \Im\left\{ r_{\acute{c}}^{\ast}\right\} \right]\Im\left\{ \tilde{u}_{m,n}\right\} }{\Re\left\{ \tilde{u}_{m,n}\right\} ^{2}+\Im\left\{ \tilde{u}_{m,n}\right\} ^{2}}.\label{eq:branch-metric-04}
\end{equation}
For CM, $\Re\left\{ \tilde{u}_{m,n}\right\} ^{2}+\Im\left\{ \tilde{u}_{m,n}\right\} ^{2}$
is constant, and hence, it can be dropped from the cost function,
which implies that no division operations are required.

To compute $J_{m,n}^{c}$, it is worth noting that the two terms in
brackets are independent of $\left\{ m,n\right\} $, and hence, they
are computed only once for each value of $c$. Therefore, the complexity
at each step in the trellis can be computed as $R_{A}=3\times2^{M}$,
$R_{M}=4+2\times2^{M}$ and $R_{D}=0$, where $2^{M}$ is the number
of branches at each step in the trellis. However, if the trellis starts
or ends by a pilot, then only $M$ computations are required. By noting
that the number of full steps is $N-2N_{P}-1$, and the number of
steps that require $M$ computations is $2\left(N_{P}-1\right)$,
then the total computations of the branch metrics (BM) are: 
\begin{align*}
R_{A}^{BM} & =\left(3\times2^{M}\right)\left(N-2N_{P}-1\right)+2\left(3\times M\right)\left(N_{P}-1\right)\\
R_{M}^{BM} & =\left(4+2^{M+1}\right)\left(N-2N_{P}-1\right)+2\left(N_{P}-1\right)\left(4+2M\right)\\
R_{D}^{BM} & =0
\end{align*}
The path metrics (PM) require $R_{A}^{PM}=\left(N-2N_{P}-1\right)+M\left(N_{P}-1\right)$
real addition. Therefore, the total complexity is: 
\begin{align}
R_{A}^{CM} & =\left(N-2N_{P}-1\right)\left(5\times2^{M}\right)+7M\left(N_{P}-1\right)\\
R_{M}^{CM} & =\left(N-2N_{P}-1\right)\left(4+2^{M+1}\right)+2\left(N_{P}-1\right)\left(4+2M\right)\\
R_{D}^{CM} & =0
\end{align}

\textcolor{black}{For QAM modulation, the most general case for the
branch metrics of the $D^{3}$ will be used as, 
\begin{equation}
J_{m,n}^{c}=\left|\frac{r_{c}}{\tilde{d}_{m}}-\frac{r_{\acute{c}}}{\tilde{d}_{n}}\right|^{2}.\label{eq:branch_metric_general_QAM}
\end{equation}
}

\textcolor{black}{The branch metric in \eqref{eq:branch_metric_general_QAM}
requires one complex addition, $C_{A}=1$, one complex multiplication,
$C_{M}=1$, and two complex divisions, $C_{D}=2$, per branch metrics.
Therefore, the total path metric complexity is: 
\begin{align}
R_{A}^{QAM} & =5MN_{P}+10M\left(N-N_{P}\right)\\
R_{M}^{QAM} & =4MN_{P}+8M\left(N-N_{P}\right)\\
R_{D}^{QAM} & =2MN_{P}+4M\left(N-N_{P}\right)
\end{align}
}

\begin{table}
\begin{onehalfspace}
\centering{}\caption{Computational complexity comparison using different values of $N$,
$N_{P}=N/4$, for BPSK.\label{tab:Computational-power-analysis}}
{\small{}}%
\begin{tabular}{|l|l|l|l|l|l|}
\hline 
{\small{}$N$} & {\small{}$128$} & {\small{}$256$} & {\small{}$512$} & {\small{}$1024$} & {\small{}$2048$}\tabularnewline
\hline 
\hline 
{\small{}$\eta_{R_{A}}$} & {\small{}$0.58$} & {\small{}$1.07$} & {\small{}$1.21$} & {\small{}$1.27$} & {\small{}$1.31$}\tabularnewline
\hline 
{\small{}$\eta_{R_{M}}$} & {\small{}$0.77$} & {\small{}$0.72$} & {\small{}$0.68$} & {\small{}$0.64$} & {\small{}$0.61$}\tabularnewline
\hline 
{\small{}$R_{D}$} & {\small{}$96$} & {\small{}$192$} & {\small{}$384$} & {\small{}$768$} & {\small{}$1536$}\tabularnewline
\hline 
{\small{}$\eta_{P}$} & {\small{}$0.20$} & {\small{}$0.21$} & {\small{}$0.22$} & {\small{}$0.26$} & {\small{}$0.31$}\tabularnewline
\hline 
\end{tabular}  \vspace{-1.5em} 
\end{onehalfspace}
\end{table}

To compare the complexity of the $D^{3}$, we use the conventional
detector using LS channel estimation, linear interpolation, zero-forcing
(ZF) equalization, and MLD, denoted as coherent-L, as a benchmark
due to its low complexity. The relative complexity is denoted by $\eta$,
which corresponds to the ratio of the $D^{3}$ complexity to the conventional
detector, i.e., $\eta_{R_{A}}$ denotes the ratio of real additions
and $\eta_{R_{M}}$ corresponds to the ratio of real multiplications.
As depicted in Table \ref{tab:Computational-power-analysis}, $R_{A}$
for $D^{3}$ less than coherent-L only using BPSK for $N=128$, and
then it becomes larger for all the other considered values of $N$.
For $R_{M}$, $D^{3}$ is always less than the coherent-L, particularly
for high values of $N$, where it becomes 0.61 for $N=2048$. It is
worth noting that $R_{D}$ in the table corresponds to the number
of divisions in the conventional OFDM since the $D^{3}$ does not
require any division operations. For a more informative comparison
between the two systems, we use the computational power analysis presented
in \cite{computational_power}, where the total power for each detector
is estimated based on the total number of operations. Table \ref{tab:Computational-power-analysis}
shows the relative computational power $\eta_{P}$, which shows that
the $D^{3}$ detector requires only $0.2$ of the power required by
the coherent-L detector for $N=128$ and $0.31\%$ for $N=2048$.

\textcolor{black}{}
\begin{table}
\begin{onehalfspace}
\centering{}\textcolor{black}{\caption{\textcolor{black}{Computational complexity comparison using different
values of $N$, $N_{P}=N/4$, for 16-QAM and 64-QAM.\label{tab:Computational-power-analysis-1}}}
}%
\begin{tabular}{|l||l|l||l|l|}
\hline 
\textcolor{black}{$M$} & \multicolumn{2}{c||}{\textcolor{black}{$16$}} & \multicolumn{2}{c|}{\textcolor{black}{$64$}}\tabularnewline
\hline 
\textcolor{black}{$N$} & \textcolor{black}{$512$} & \textcolor{black}{$2048$} & \textcolor{black}{$512$} & \textcolor{black}{$2048$}\tabularnewline
\hline 
\hline 
\textcolor{black}{$\eta_{R_{A}}$} & \textcolor{black}{$1.25$} & \textcolor{black}{$1.25$} & \textcolor{black}{$1.64$} & \textcolor{black}{$1.64$}\tabularnewline
\hline 
\textcolor{black}{$\eta_{R_{M}}$} & \textcolor{black}{$0.52$} & \textcolor{black}{$0.47$} & \textcolor{black}{$0.64$} & \textcolor{black}{$0.62$}\tabularnewline
\hline 
\textcolor{black}{$\eta_{R_{D}}$} & \textcolor{black}{$0.98$} & \textcolor{black}{$0.98$} & \textcolor{black}{$0.99$} & \textcolor{black}{$0.99$}\tabularnewline
\hline 
\textcolor{black}{$\eta_{P}$} & \textcolor{black}{$0.94$} & \textcolor{black}{$0.84$} & \textcolor{black}{$0.91$} & \textcolor{black}{$0.80$}\tabularnewline
\hline 
\end{tabular}\textcolor{black}{  \vspace{-1.5em} }
\end{onehalfspace}
\end{table}

\textcolor{black}{It is also worth considering the complexity analysis
for higher modulation orders that require division operations such
as 16-QAM and 64-QAM since they widely used in modern wireless broadband
systems \cite{WiMax}, \cite{LTE-A}. Table \ref{tab:Computational-power-analysis-1}
shows the rations of real multiplications, multiplications, divisions,
and lastly the ration of the overall computational power for 16-QAM
and 64-QAM considering $N=512$ and $N=2048$. Unlike the CM modulus
case, the $D^{3}$ requires division operations, where it is very
comparable to conventional OFDM receivers in terms of the division
computational resources. Although, the total number of computational
addition resources needed is higher in $D^{3}$ by $25\%-65\%$, Nevertheless,
the overall computational resources in $D^{3}$ is less than the conventional
OFDM reveries by $\%6-20\%$ due to the significant saving in the
multiplication operations of the $D^{3}$.}

Besides, it is worth noting that linear interpolation has lower complexity
as compared to more accurate interpolation schemes such as the spline
interpolation \cite{spline-interpolation}, \cite{Spline}, which
comes at the expense of the error rate performance. Therefore, the
results presented in Table \ref{tab:Computational-power-analysis}
can be generally considered as upper bounds on the relative complexity
of the $D^{3}$, when more accurate interpolation schemes are used,
the relative complexity will drop even further as compared to the
results in Table \ref{tab:Computational-power-analysis}.

\subsection{Complexity with Error Correction Coding}

To evaluate the impact of the complexity reduction of the $D^{3}$
in the presence of FEC coding, convolutional codes are considered
with soft and hard decision decoding using the VA. BPSK is the modulation
considered for the complexity evaluation and the code rate is assumed
to be $1/2$. For decoding of convolutional codes, the soft VA requires
$n\times2^{K}$ addition or subtractions and multiplications per decoded
bit, where $1/n$ is the code rate and $K$ is the constraint length
\cite{P-Wu}. Therefore, for $1/2$ code rate, $R_{A}=R_{M}=2^{K+1}$.
Given that each OFDM symbol has $N$ coded bits and $N/2$ information
bits, the complexity per OFDM symbol becomes $R_{A}=R_{M}=N\times2^{K}$.
For the hard VA, $N\times2^{K}$ XOR operations are required for the
branch metric computation, while $N\times2^{K-1}$ additions are required
for the path metric computations. Because the XOR operation is a bit
operation, it's complexity is much less than the addition. Assuming
that addition is using an 8-bit representation, then the complexity
of an addition operation is about eight times the XOR. Therefore,
$R_{A}$, in this case, can be approximated as $N\left(2^{K}+2^{K-2}\right)$.

\begin{table}
\begin{onehalfspace}
\centering{}\caption{Computational complexity comparison using hard and soft VA for different
values of $K$, $N=2048$. \label{T-coded}}
{\small{}}%
\begin{tabular}{|c|c|c|c|c|c|}
\hline 
{\small{}$K$ } & {\small{}$3$ } & {\small{}$4$ } & {\small{}$5$} & {\small{}$6$} & {\small{}$7$}\tabularnewline
\hline 
\hline 
\multicolumn{1}{|l|}{\textbf{\small{}Soft}} & {\small{}$0.96$ } & {\small{}$0.97$ } & {\small{}$0.97$ } & {\small{}$0.98$ } & {\small{}$0.99$ }\tabularnewline
\hline 
\multicolumn{1}{|l|}{\textbf{\small{}Hard}} & {\small{}$0.24$ } & {\small{}$0.26$ } & {\small{}$0.28$ } & {\small{}$0.33$ } & {\small{}$0.41$ }\tabularnewline
\hline 
\end{tabular}  \vspace{-1.5em} 
\end{onehalfspace}
\end{table}

As can be noted from Table \ref{T-coded}, the complexity reduction
when soft VA is used less significant as compared to the hard VA.
Such a result is obtained because the soft VA requires the CSI to
compute the reliability factors, which requires $N-N_{P}$ division
operations when the $D^{3}$ is used. For hard decoding, the advantage
of the $D^{3}$ is significant even for high constraint length values.

\section{Numerical Results\label{sec:Numerical-Results}}

This section presents the performance of the $D^{3}$ detector in
terms of BER for several operating scenarios. The system model follows
the LTE-A physical layer (PHY) specifications \cite{LTE-A}, where
the adopted OFDM symbol has $N=512$, $N_{\mathrm{CP}}=64$, the sampling
frequency $f_{s}=7.68$ MHz, the subcarrier spacing $\Delta f=15$
kHz, and the pilot grid follows that of Fig. \ref{fig:2D-to-1D}.
The total OFDM symbol period is $75$ $\mu\sec$, and the CP period
is $4.69$ $\mu\sec$. The channel models used are the flat Rayleigh
fading channel, the typical urban (TUx) multipath fading model \cite{Typical Urban}
that consists of $6$ taps with normalized delays of $\left[0,2,3,9,13,29\right]$
and average taps gains are $\left[0.2,0.398,0.2,0.1,0.063,0.039\right]$,
which corresponds to a severe frequency-selective channel. The TUx
model is also used to model a moderate frequency-selective channel
where the number of taps in the channel is $9$ with normalized delays
of $[0$, $1$, $\ldots$, $8]$ samples, and the average taps gains
are $[0.269$, $0.174$, $0.289$, $0.117$, $0.023$, $0.058$, $0.036$,
$0.026$, $0.008]$. The channel taps gains are assumed to be independent
and Rayleigh distributed. The Monte Carlo simulation results included
in this work are obtained by generating $10^{6}$ OFDM symbols per
simulation run. Throughout this section, the ML coherent detector
with perfect CSI will be denoted as coherent, while the coherent with
linear and spline interpolation will be denoted as coherent-L and
coherent-S, respectively. Moreover, the results are presented for
the SISO system, $\mathit{\mathcal{N}\mathrm{=1}}$, unless it is
mentioned otherwise. \textcolor{black}{The SNR in the obtained results
is defined as the ratio of the average received signal power to the
average noise power regardless of the number of pilots. Such an approach
is followed because the proposed system in this work is evaluated
in the context of the LTE RB, which has a fixed structure. For more
general comparisons, the power and spectral efficiency of all considered
systems should be identical.}

\begin{figure}
\begin{minipage}[t]{0.48\columnwidth}%
\begin{center}
\includegraphics[width=0.4\paperwidth]{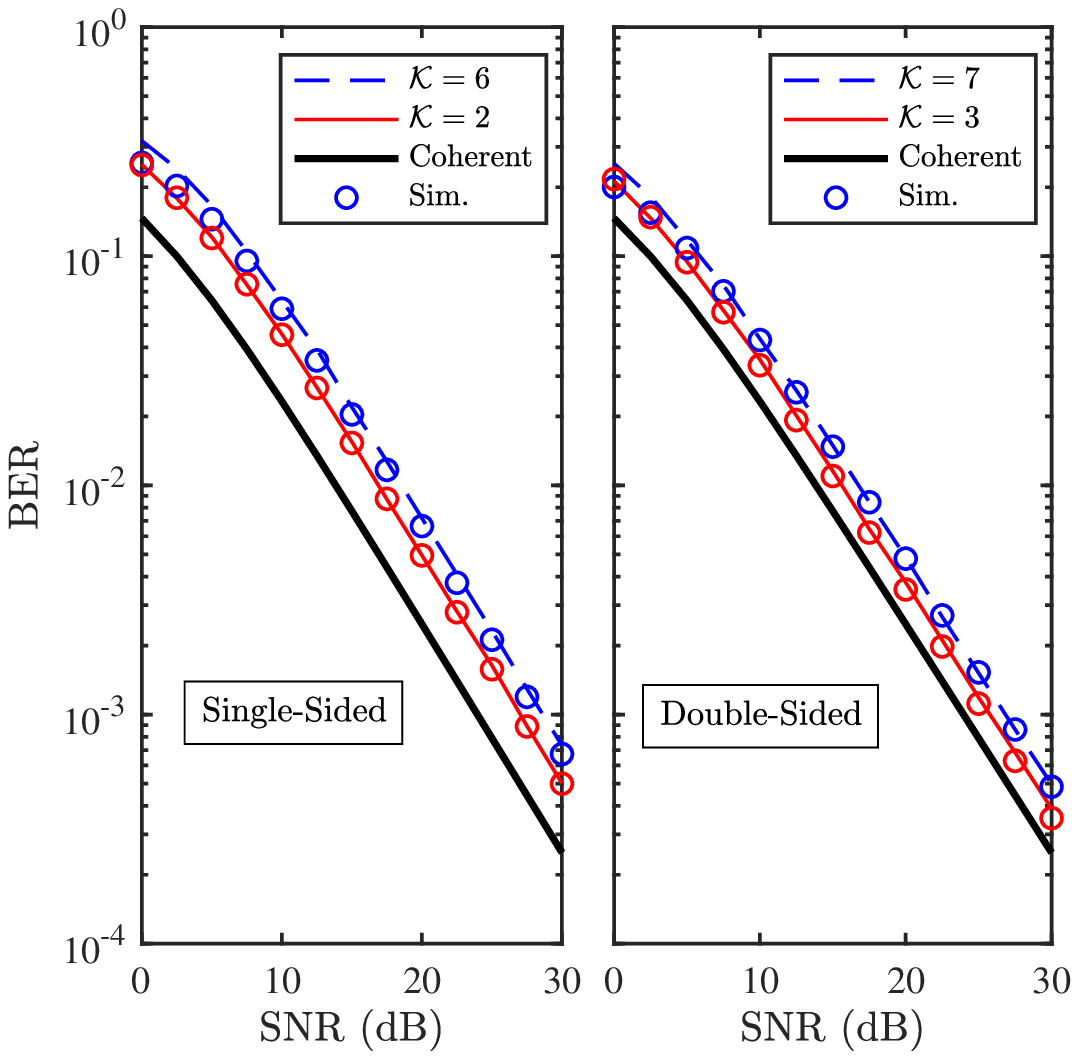}
\par\end{center}
\begin{singlespace}
\caption{{\footnotesize{}BER using SS and DS pilots for different values of
$\mathcal{K}$ over flat fading channels using BPSK, $\mathcal{N}=1.$\label{fig:BER-Single-Double-Sided-Flat}}}
\end{singlespace}
\end{minipage}\hfill{}%
\begin{minipage}[t]{0.48\columnwidth}%
\begin{center}
\includegraphics[width=0.4\paperwidth]{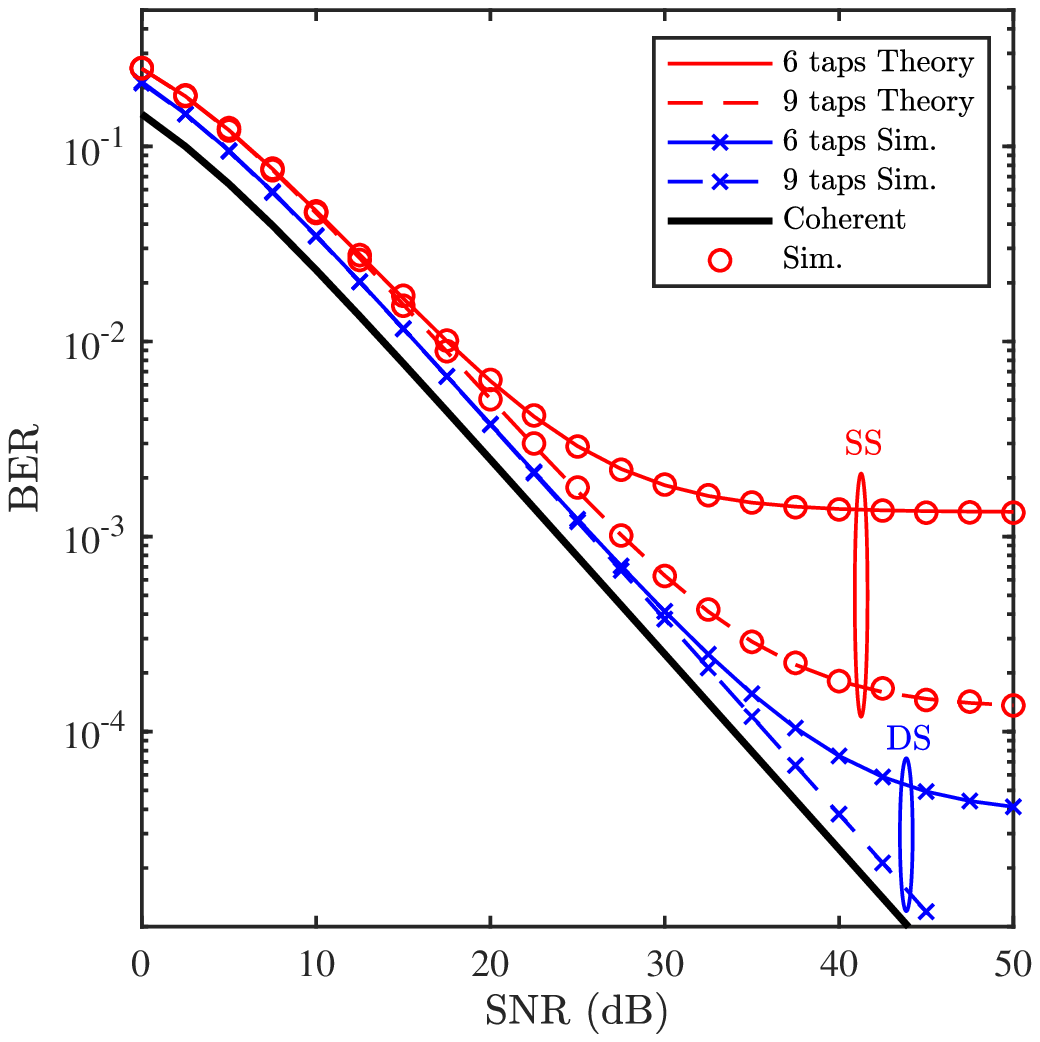}
\par\end{center}
\begin{singlespace}
\caption{BER in frequency-selective channels using BPSK, $\mathcal{K}_{D}=1$
and $\mathcal{N}=1$. \label{fig:BER-SISO-D3-SS-6-taps} }
  \vspace{-1.5em} 
\end{singlespace}
\end{minipage}
\end{figure}

Fig. \ref{fig:BER-Single-Double-Sided-Flat} shows the BER of the
single-sided (SS) and double-sided (DS) $D^{3}$ over flat fading
channels for $\mathcal{K}=2,6$ and $3,7$, respectively, and using
BPSK. The number of data symbols $\mathcal{K}_{D}=\mathcal{K}-1$
for the SS and $\mathcal{K}_{D}=\mathcal{K}-2$ for the DS because
there are two pilot symbols at both ends of the data segment for the
DS case. The results in the figure for the SS show that $\mathcal{K}$
has a noticeable impact on the BER where the difference between the
$\mathcal{K}=2$ and $6$ cases is about $1.6$ dB at BER of $10^{-3}$.
For the DS segment, the BER has the same trends of the SS except that
it becomes closer to the coherent case because having more pilots
reduces the probability of sequence inversion due to the phase ambiguity
problem. The figure shows that the approximated and simulation results
match very well for all cases, which confirms the accuracy of the
derived approximations.

The effect of the frequency selectivity is illustrated in Fig. \ref{fig:BER-SISO-D3-SS-6-taps}
for the SS and DS configurations using$\mathcal{K}_{D}=1$. As can
be noted from the figure, frequency-selective channels introduce error
floors at high SNRs, which is due to the difference between adjacent
channel values caused by the channel frequency selectivity. Furthermore,
the figure shows a close match between the simulation and the derived
approximations. The approximation results are presented only for $\mathcal{K}=2$
because evaluating the BER for $\mathcal{K}>2$ becomes computationally
prohibitive. For example, evaluating the integral \eqref{eq:unconditional-SER}
for the $\mathcal{K}=3$ requires solving a $6$-fold integral. The
results for the frequency-selective channels are quite different from
the flat fading cases. In particular, the BER performance drastically
changes when the DS pilot segment is used. Moreover, the impact of
the frequency selectivity is significant, particularly for the SS
pilot case.

\begin{figure}
\begin{minipage}[t]{0.48\columnwidth}%
\begin{center}
\includegraphics[width=0.4\paperwidth]{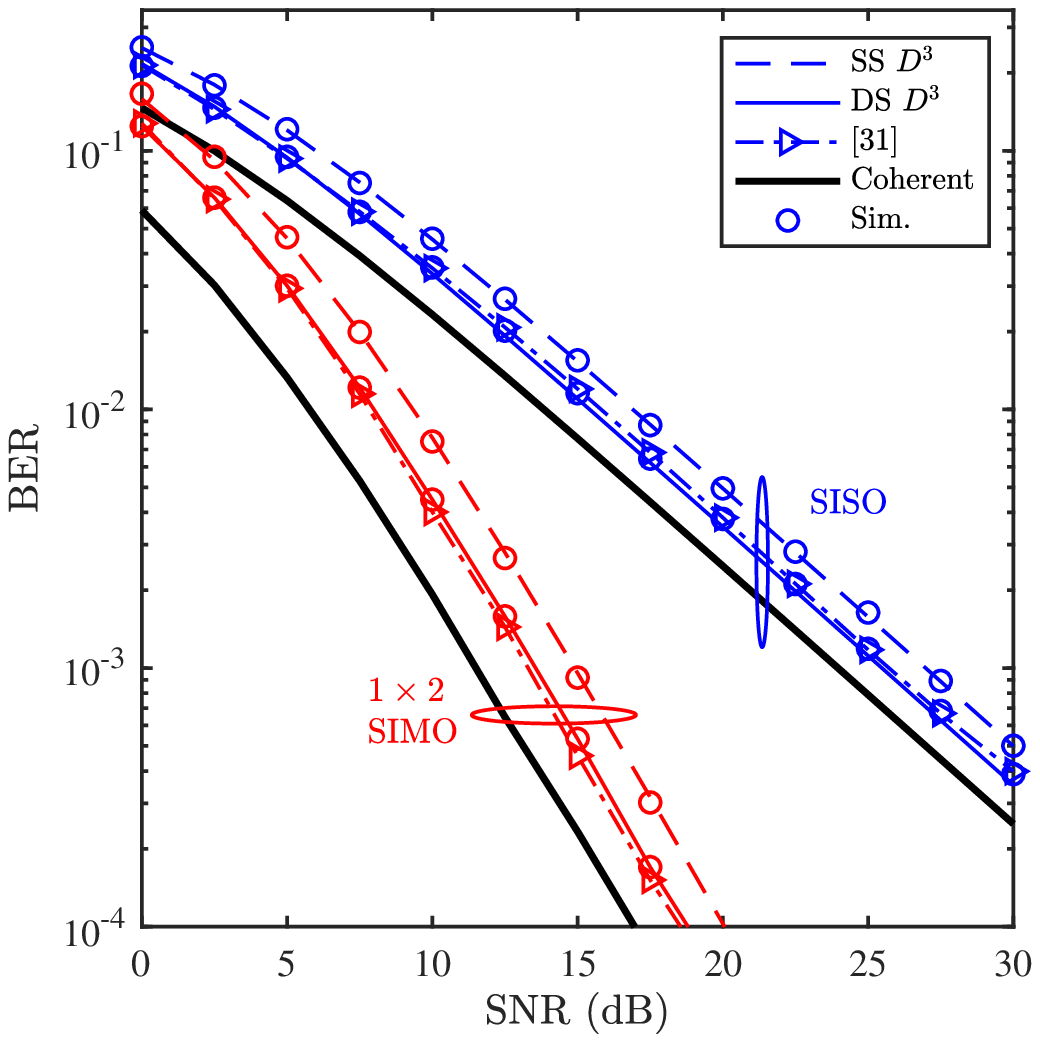}
\par\end{center}
\caption{BER of $D^{3}$ and MLSD \cite{Wu 2010} SISO and SIMO using SS and
DS pilots, flat fading, BPSK, $\mathcal{N}=1$, $2$, and $\mathcal{K}_{D}=1.$\label{fig:BER-SIMO-D3-flat} }
  \vspace{-1.5em} %
\end{minipage}\hfill{}%
\begin{minipage}[t]{0.48\columnwidth}%
\begin{center}
\includegraphics[width=0.4\paperwidth]{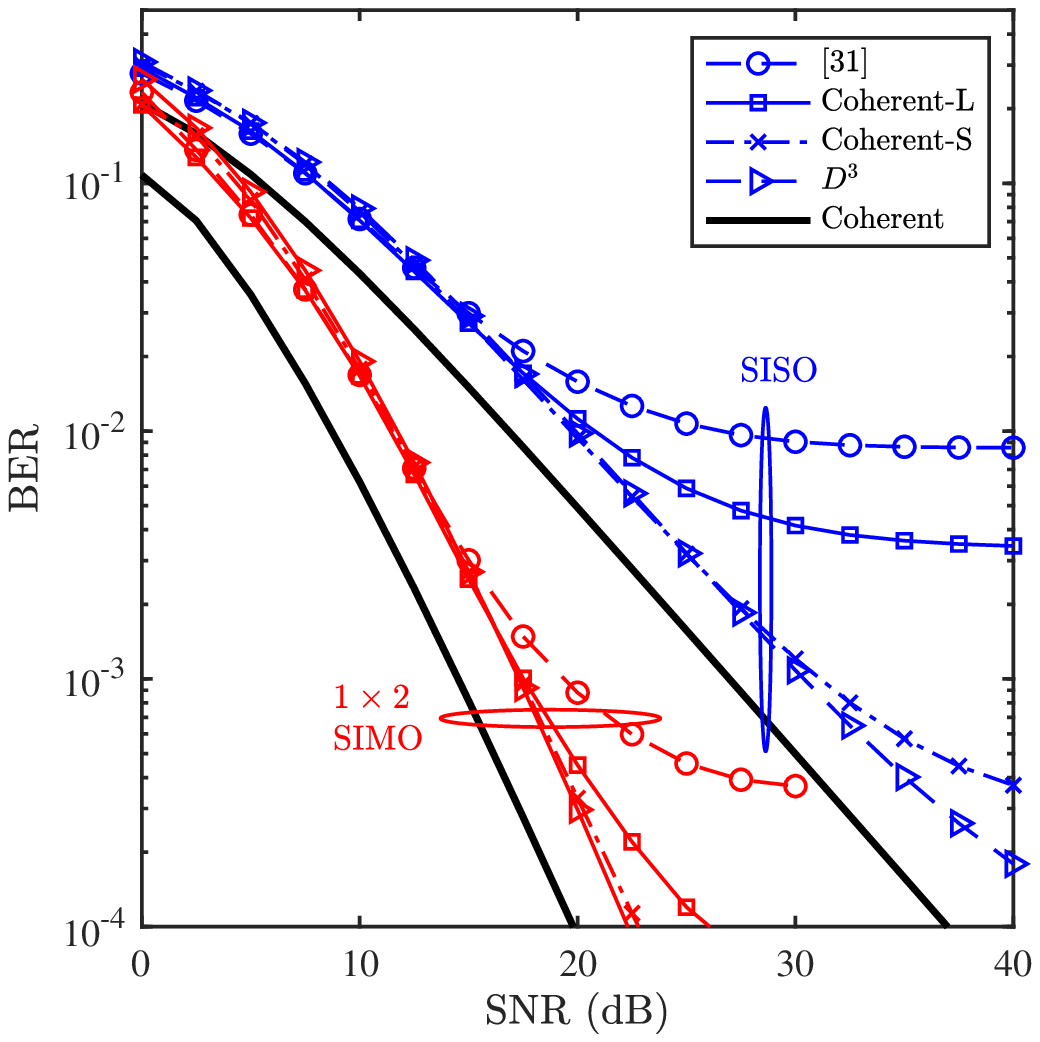}
\par\end{center}
\caption{BER of the SISO $D^{3}$ and MLSD \cite{Wu 2010} over the 6-taps
frequency-selective channel using QPSK, $\mathcal{K}_{D}=1$, $\mathcal{N}=1$,
$2.$\label{fig:BER-SISO-D3-QPSK} }

  \vspace{-1.5em} %
\end{minipage}
\end{figure}

Fig. \ref{fig:BER-SIMO-D3-flat} shows the BER of the $1\times2$
SIMO $D^{3}$ over flat fading channels for SS and DS pilot segments.
It can be noted from the figure that the maximum ratio combiner (MRC)
BER with perfect CSI outperforms the DS and SS systems by about $2$
and $3$ dB, respectively. Moreover, the figure shows that the MLSD
\cite{Wu 2010} and the $D^{3}$ have equivalent BER for the SISO
and SIMO scenarios. \textcolor{black}{The figure also shows the BER
of the 1\texttimes 2 SIMO systems as compared to the SISO case. }

Figs. \ref{fig:BER-SISO-D3-QPSK} shows the BER of the SISO and $1\times2$
SIMO MLSD, coherent, coherent-S and coherent-L systems over frequency-selective
channels. For both SISO and SIMO, the BER of all the considered techniques
converges at low SNRs because the AWGN dominates the BER in the low
SNR range. For moderate and high SNRs, the $D^{3}$ outperforms all
the other considered techniques except for the coherent, where the
difference is about $3.5$ and $2.75$ dB at BER of $10^{-3}$ for
the SISO and SIMO\ systems, respectively.

\begin{figure}
\begin{minipage}[t]{0.48\columnwidth}%
\begin{center}
\includegraphics[width=0.4\paperwidth]{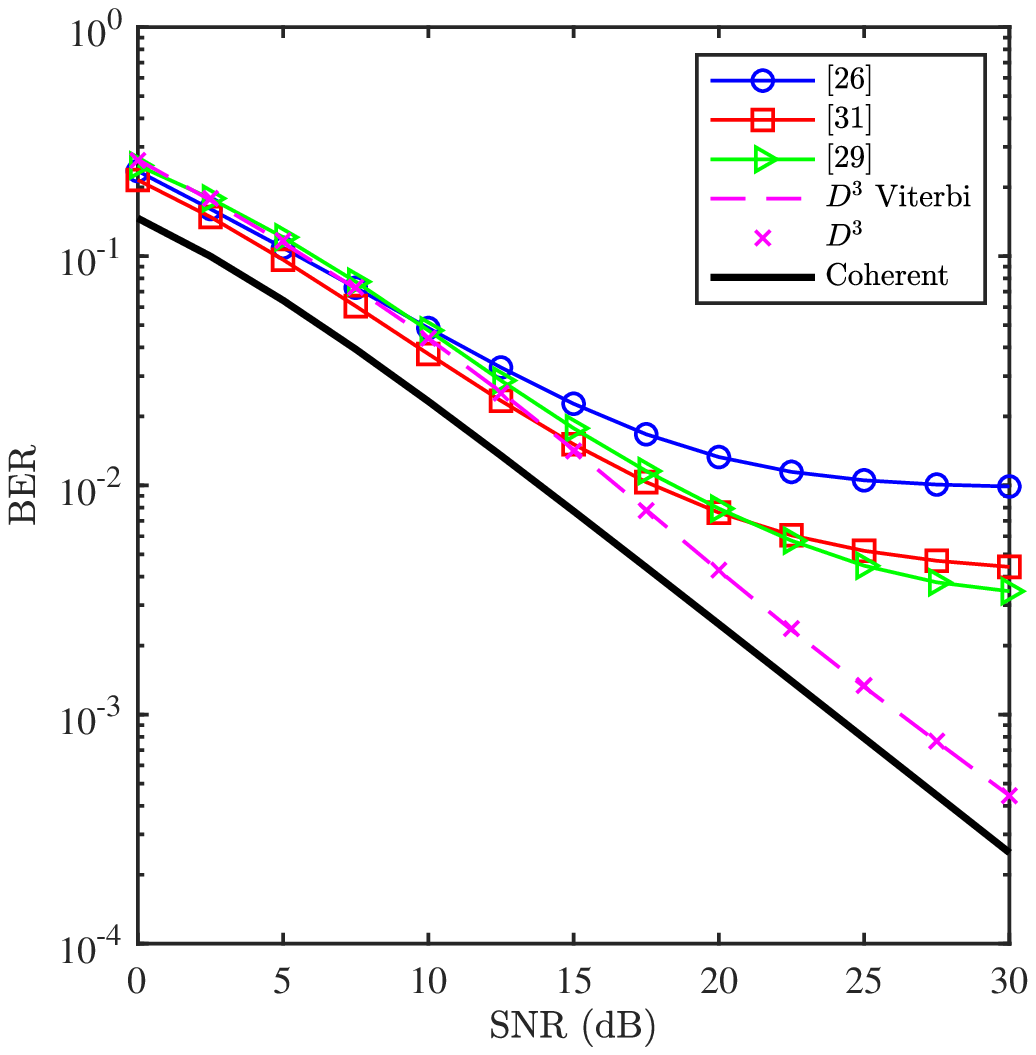}
\par\end{center}
\caption{\textcolor{black}{BER of the $D^{3}$ for $\mathcal{K}=7$ DS using
BPSK compared to several other sequence detectors over 6-taps frequency-selective
channel.}\textcolor{red}{\label{fig:BER-comparasion-psp} }}

  \vspace{-1.5em} %
\end{minipage}\hfill{}%
\begin{minipage}[t]{0.48\columnwidth}%
\begin{center}
\includegraphics[width=0.4\paperwidth]{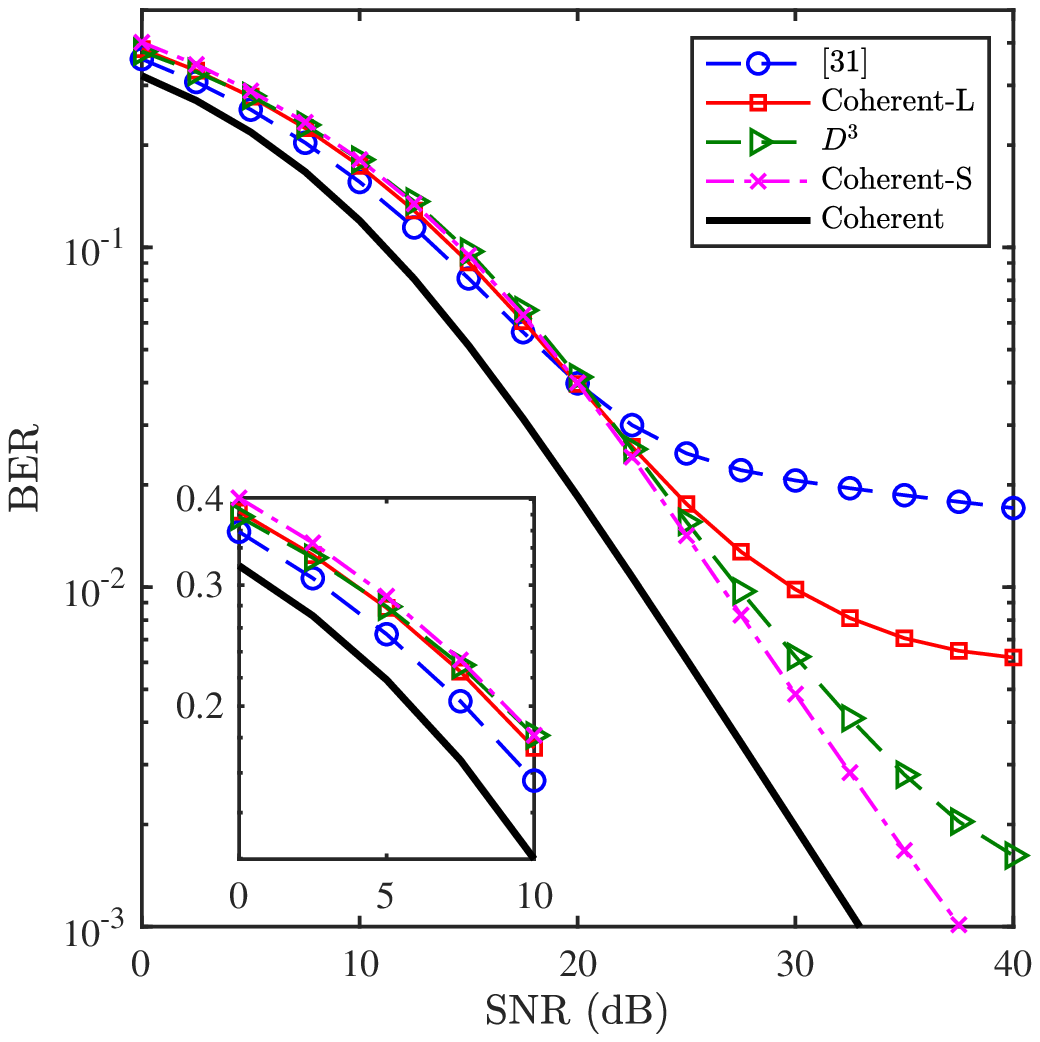}
\par\end{center}
\begin{singlespace}
\caption{\textcolor{black}{BER of the $D^{3}$ for $\mathcal{K}=7$ DS over
the 6-taps frequency-selective channel using 16-QAM, compared with
MLSD \cite{Wu 2010}.}\textcolor{brown}{\label{fig:BER-SISO-D3-16QAM}} }
  \vspace{-1.5em} 
\end{singlespace}
\end{minipage}
\end{figure}

\textcolor{black}{Fig. \ref{fig:BER-comparasion-psp} compares the
BER of the $D^{3}$, PSP \cite{PSP-Raheli}, MLSD \cite{Wu 2010},
MSDD \cite{Divsalar}, and the coherent detector over the 6-taps channel
using BPSK. As can be noted from the figure, the $D^{3}$ noticeably
outperforms all other detectors for $\mathrm{SNR}\gtrsim15$ dB, which
indicates that the $D^{3}$ is more robust to the frequency selectivity
of the channel. Moreover, the figure shows the $\mathit{D^{\mathrm{3}}}$BER
using VA which, as expected, is identical to the BER obtained using
(\ref{E-DDD-00}). It is worth noting that all the systems considered
in the figure are implemented using the DS segment where $\mathcal{K}=7$,
and thus, they are evaluated under similar throughput conditions.
However, the BER sensitivity of each technique to the number of pilot
symbols could be different from other techniques, which implies that
some of these techniques might be able to provide roughly the same
BER but using fewer pilot symbols. The same argument applies to the
power efficiency as well, because the power allocated per information
bit becomes different for various systems. However, because the LTE
RB is used as the basis for testing all systems, then the current
comparison can be considered generally fair. In the worst case scenario,
i.e., considering that all other systems are fully blind, then the
throughput power loss is only 4.7\% as described in Subsection \ref{subsec:Resource-Block-Detection},
which has a negligible effect on the BER.}

\textcolor{black}{Fig. \ref{fig:BER-SISO-D3-16QAM} shows the BER
for the $D^{3}$, MLSD \cite{Wu 2010}, coherent, coherent-L and coherent-S
using $16$-QAM. As can be noted from the figure, the MLSD slightly
outperforms the $D^{3}$ at low SNRs, and the coherent-S outperforms
the $D^{3}$ at high SNRs. However, the coherent-S has generally much
higher complexity.}

\begin{figure}
\begin{minipage}[t]{0.48\columnwidth}%
\begin{center}
\includegraphics[width=0.4\paperwidth]{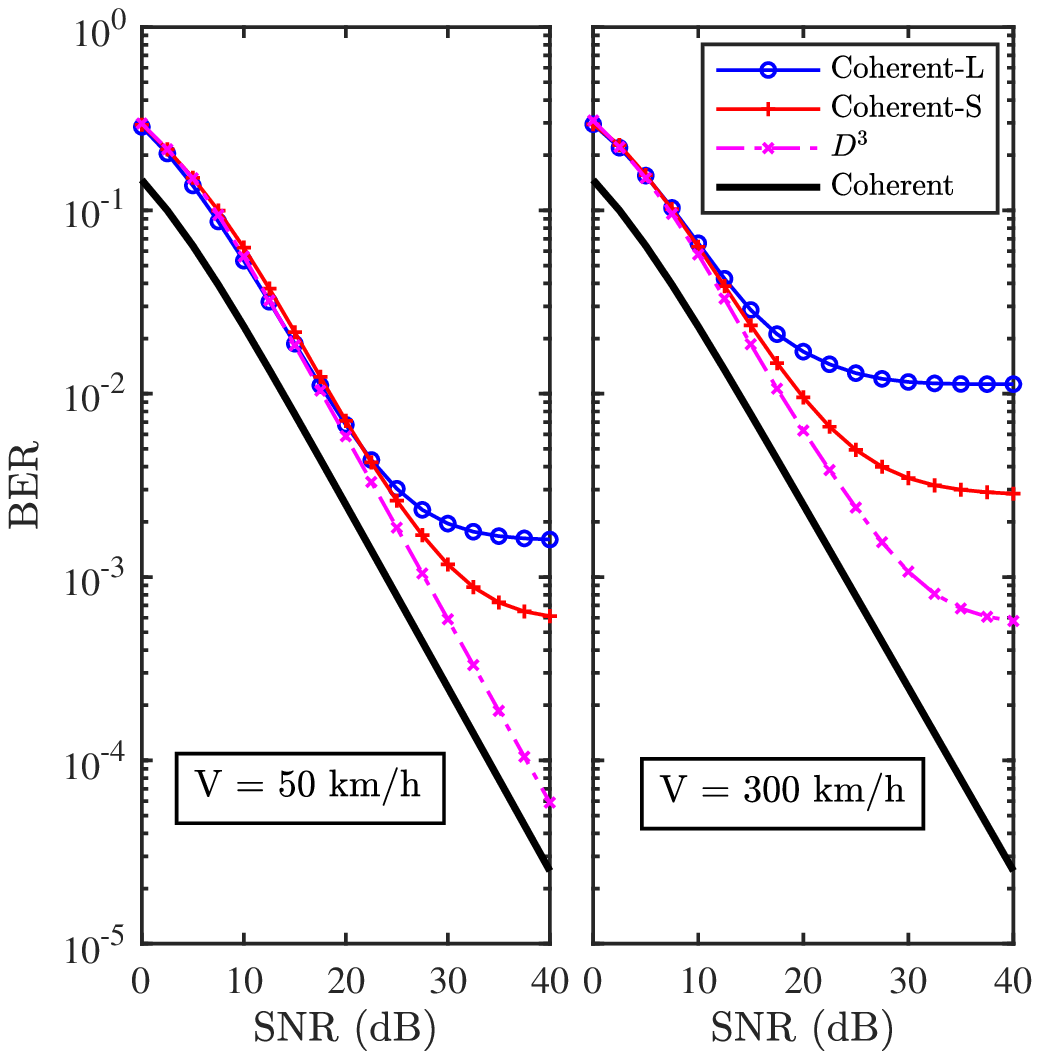}
\par\end{center}
\begin{singlespace}
\caption{\textcolor{black}{BER for the SISO $D^{3}$, coherent-L, and coherent
detector for a complete LTE RB using the 6-taps channel and BPSK for
different mobility values. }\textcolor{blue}{\label{fig:BER-D3-Full-RB}}}
  \vspace{-1.5em} 
\end{singlespace}
\end{minipage}\hfill{}%
\begin{minipage}[t]{0.48\columnwidth}%
\begin{center}
\includegraphics[width=0.4\paperwidth]{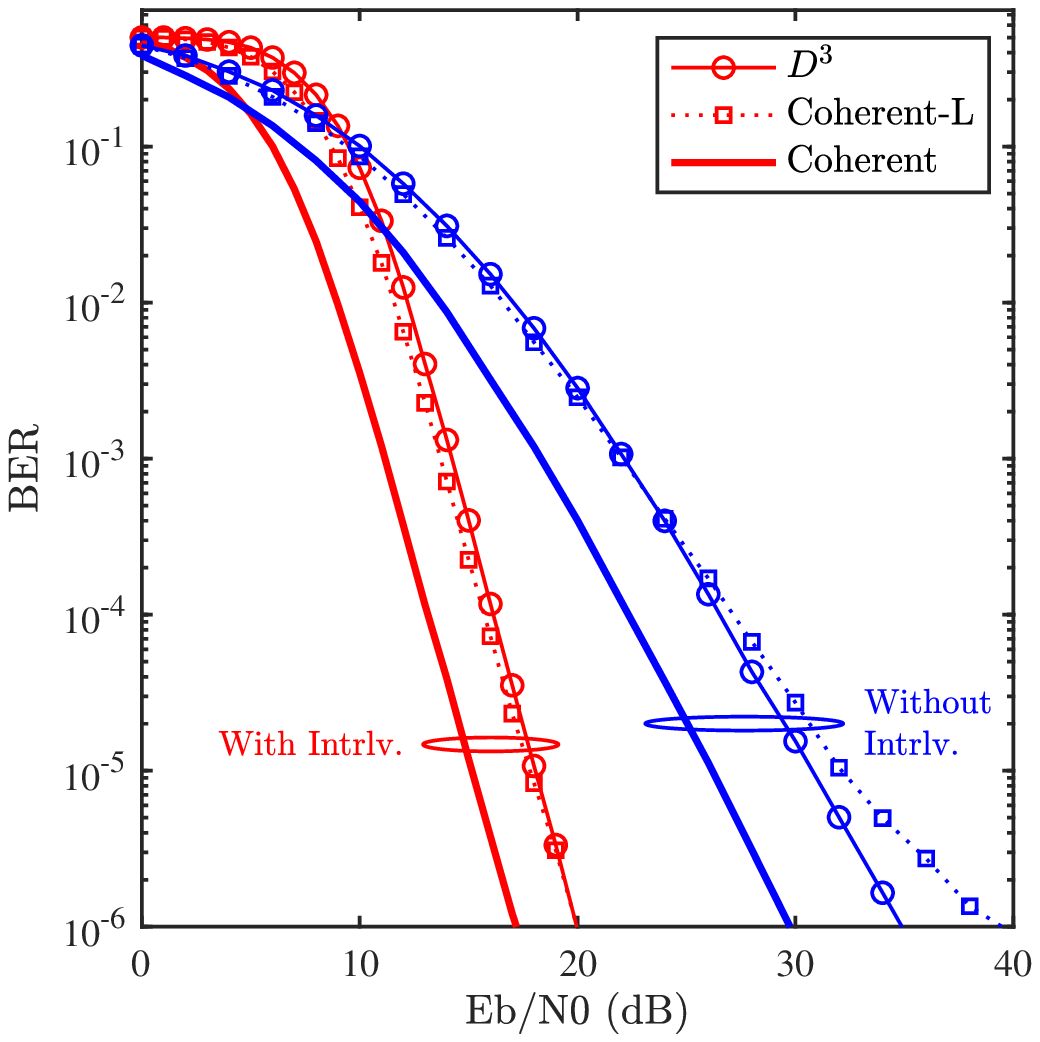}
\par\end{center}
\caption{\textcolor{black}{Coded BER for the SISO $D^{3}$, coherent-L, and
coherent detector for $\mathcal{K}=7$ DS over the 6-taps frequency-selective
channel using BPSK. }\textcolor{blue}{\label{fig:Coded-BER}}}

  \vspace{-1.5em} %
\end{minipage}
\end{figure}

\textcolor{black}{Fig. \ref{fig:BER-D3-Full-RB} shows the simulated
BER of the $D^{3}$ system when it is used to detect a complete RB
as described in Subsection \ref{subsec:Resource-Block-Detection}.
The channel model is similar to the 6-taps used described above, and
the channel gain variation over consecutive OFDM symbols is generated
using the Jakes's model, where the maximum Doppler frequency $f_{d}=\frac{V}{c}\,f_{c}$,
where $V$ is the speed of the vehicle, $c$ is the speed of light,
$c=3\times10^{8}$ m/s, and the carrier $f_{c}=1.9$ GHz. The channel
is considered quasi-static, i.e., the channel remains constant over
the OFDM symbol period, but changes over consecutive symbols. As the
figure indicates, the $D^{3}$ is more immune to channel mobility
at $50$ km/h as compared to pilot-based systems as it did not have
an error floor. For the high mobility case, $V=300$ km/h, the $D^{3}$
BER exhibited an error floor at about $6\times10^{-4}$, which is
much lower than the error floor of the coherent detector with linear
and spline interpolation.}

\textcolor{black}{Fig. \ref{fig:Coded-BER} shows the simulated BER
of the $D^{3}$ using convolutional codes with hard decision decoding,
using the widely used ($171$, $131$) convolutional code with a block
length of $256$ bits, and a $512\times512$ channel block interleaver.
Moreover, the results without interleaver are considered, which corresponds
to the case of slow fading channels with very long coherence time.
As it can be noted from the figure, the BER of the $D^{3}$ and coherent-L
are comparable for the considered range of SNR when the block interleaver
is used. On the contrary, with no interleaving, the $D^{3}$ offers
about $5$ dB advantage at $10^{-6}$. Both detectors are approximately
$3$ dB away from the coherent detector with perfect CSI.}

\section{Conclusion and Future Work\label{sec:Conclusion}}

This work proposed a new receiver design for OFDM-based broadband
communication systems. The new receiver performs the detection process
directly from the FFT output symbols without the need of experiencing
the conventional steps of channel estimation, interpolation, and equalization,
which led to a considerable complexity reduction. Moreover, the $D^{3}$
system can be deployed efficiently using the VA. The proposed system
was analyzed theoretically where simple closed-form expressions were
derived for the BER in several cases of interest. The analytical and
simulation results show that the $D^{3}$ BER outperforms the coherent
pilot-based receiver in various channel conditions, particularly in
frequency-selective channels where the $D^{3}$ demonstrated high
robustness.

Although the $D^{3}$ may perform well even in severe fading conditions,
it is crucial to evaluate its sensitivity to various practical imperfections.
Thus, we will consider in our future work the performance of the $D^{3}$
in the presence of various system imperfections such as phase noise,
synchronization errors and IQ imbalance. Moreover, we will evaluate
the $D^{3}$ performance in mobile fading channels, where the channel
variation may introduce intercarrier interference.

\section*{Appendix I}

By defining the events $A_{\psi}>A_{n}\triangleq E_{\psi,n}$, $n\in\left\{ 0\text{, }1\text{, }\ldots,\psi-1\right\} $,
then, 
\begin{equation}
P_{C}|_{\mathbf{H}_{0},\mathbf{\mathbf{1}}}=P\left(\bigcap\limits _{n=0}^{\psi-1}E_{\psi,n}\right).\label{E-PC-01}
\end{equation}
Using the chain rule, $P_{C}|_{\mathbf{H}_{0},\mathbf{\mathbf{1}}}$
can be written as, 
\begin{equation}
P_{C}|_{\mathbf{H}_{0},\mathbf{\mathbf{1}}}=\Pr\left(\left.E_{\psi,\psi-1}\right\vert \bigcap\limits _{n=0}^{\psi-2}E_{\psi,n}\right)\Pr\left(\bigcap\limits _{n=0}^{\psi-2}E_{\psi,n}\right).
\end{equation}
For $\mathcal{K}=2$, $\psi=1$, $\tilde{\mathbf{d}}_{0}^{(0)}=[1$,
$-1]$, $\tilde{\mathbf{d}}_{0}^{(1)}=[1$,$1]$, and thus, 
\begin{eqnarray}
P_{C}|_{\mathbf{H}_{0},\mathbf{\mathbf{1}}} & = & \Pr\left(E_{1,0}\right)\nonumber \\
 & = & \Pr\left(\Re\left\{ r_{1}r_{2}\right\} >\Re\left\{ -r_{1}r_{2}\right\} \right)=\Pr\left(\Re\left\{ r_{0}r_{1}\right\} >0\right).
\end{eqnarray}
For $\mathcal{K}=3$, $\psi=4$, $\tilde{\mathbf{d}}_{0}^{(0)}=[1$,
$1$, $-1]$, $\tilde{\mathbf{d}}_{0}^{(1)}=[1$, $-1$, $-1]$, $\tilde{\mathbf{d}}_{0}^{(2)}=[1$,
$-1$, $1]$ and $\tilde{\mathbf{d}}_{0}^{(3)}=[1$, $1$,\ldots$,1]$ .
Using the chain rule 
\begin{eqnarray}
P_{C}|_{\mathbf{H}_{0},\mathbf{\mathbf{1}}} & = & \Pr\left(E_{3,2}|E_{3,1}\text{, }E_{3,0}\right)\Pr\left(E_{3,1},E_{3,0}\right)\nonumber \\
 & = & \Pr\left(E_{3,2}|E_{3,1}\text{, }E_{3,0}\right)\Pr\left(E_{3,1}|E_{3,0}\right)\Pr\left(E_{3,0}\right)\label{E-PrA3A0}
\end{eqnarray}
However, $\Pr\left(E_{3,0}\right)=\Pr\left(A_{3}>A_{0}\right)$, and
thus 
\begin{eqnarray}
\Pr\left(E_{3,0}\right) & = & \Pr\left(\Re\left\{ r_{0}r_{1}+r_{1}r_{2}\right\} >\Re\left\{ r_{0}r_{1}-r_{1}r_{2}\right\} \right)\nonumber \\
 & = & \Pr\left(\Re\left\{ r_{1}r_{2}\right\} >\Re\left\{ -r_{1}r_{2}\right\} \right)=\Pr\left(\Re\left\{ r_{1}r_{2}\right\} >0\right).
\end{eqnarray}
The second term in (\ref{E-PrA3A0}) can be evaluated by noting that
the events $E_{3,1}$ and $E_{3,0}$ are independent. Therefore $\Pr\left(E_{3,1}|E_{3,0}\right)=\Pr\left(E_{3,1}\right)$,
which can be computed as 
\begin{eqnarray}
\Pr\left(E_{3,1}\right) & = & \Pr\left(\Re\left\{ r_{0}r_{1}+r_{1}r_{2}\right\} >\Re\left\{ -r_{0}r_{1}+r_{1}r_{2}\right\} \right)\nonumber \\
 & = & \Pr\left(\Re\left\{ r_{0}r_{1}\right\} >\Re\left\{ -r_{0}r_{1}\right\} \right)=\Pr\left(\Re\left\{ r_{0}r_{1}\right\} >0\right).
\end{eqnarray}
The first term in (\ref{E-PrA3A0}) $\Pr\left(E_{3,2}|E_{3,1}\text{, }E_{3,0}\right)=1$
because if $A_{3}>\left\{ A_{1},A_{0}\right\} $, then $A_{3}>A_{2}$
as well. Consequently, 
\begin{equation}
P_{C}|_{\mathbf{H}_{0},\mathbf{\mathbf{1}}}=\Pr\left(\Re\left\{ r_{0}r_{1}\right\} >0\right)\Pr\left(\Re\left\{ r_{1}r_{2}\right\} >0\right).
\end{equation}
By induction, it is straightforward to show that $P_{C}|_{\mathbf{H}_{0},\mathbf{\mathbf{1}}}$
can be written as, 
\begin{equation}
P_{C}\mid_{\mathbf{H},\mathbf{\mathbf{d}=\mathbf{1}}}=\prod\limits _{n=0}^{\mathcal{K-}2}\Pr\left(\Re\left\{ r_{n}r_{\acute{n}}\right\} >0\right).
\end{equation}

\end{document}